\documentclass[aps,prx,twocolumn,amsmath,amssymb,superscriptaddress,longbibliography]{revtex4-1}

\usepackage{graphicx}
\usepackage{bm}
\usepackage{graphics}
\usepackage{subfigure}
\usepackage{graphicx}
\usepackage{epsfig}
\usepackage{epstopdf}
\usepackage{slashed}
\usepackage{color}
\usepackage[normalem]{ulem}
\usepackage[titletoc]{appendix}
\usepackage{array}

\allowdisplaybreaks

\usepackage{hyperref}
\hypersetup{colorlinks = true, urlcolor = blue, linkcolor = blue, citecolor = blue}
\newcolumntype{C}{>{$\displaystyle} c <{$}}
\newcolumntype{M}[1]{>{$\displaystyle\qquad}p{#1}<{$}}

\newcommand{\kF}{k_{\mathrm{F}}}
\newcommand{\EF}{\varepsilon_{\mathrm{F}}}
\newcommand{\EFLR}{\varepsilon_{\mathrm{F},L/R}}

\newcommand{\kFLR}{k_{\mathrm{F},L/R}}

\newcommand{\qkj}{q_{\kappa,j}}

\newcommand{\vFermikj}{\tilde{v}_{\kappa,j}}
\newcommand{\vFermipj}{\tilde{v}_{p,j}}
\newcommand{\Psikj}{\Psi_{\kappa,j}}
\newcommand{\Idc}{I_{\mathrm{dc}}}

\newcommand{\akj}{\hat{a}_{\kappa,j}}

\newcommand{\qpL}{q_{pL}}
\newcommand{\qhL}{q_{hL}}
\newcommand{\qhR}{q_{hR}}
\newcommand{\qpR}{q_{pR}}

\newcommand{\khR}{k_{hR}}
\newcommand{\kpR}{k_{pR}}
\newcommand{\uL}{u_{L}}
\newcommand{\vL}{v_{L}}
\newcommand{\uR}{u_{R}}
\newcommand{\vR}{v_{R}}
\newcommand{\NLnp}{N_{L,n}^p}
\newcommand{\NLnh}{N_{L,n}^h}
\newcommand{\ALn}{A_{L,n}}
\newcommand{\ARnp}{A_{R,n+1}}

\newcommand{\ARnm}{A_{R,n-1}}
\newcommand{\JLnp}{J_{L,n}^p}

\newcommand{\JLp}{J_{L}^p}
\newcommand{\JLh}{J_{L}^h}

\newcommand{\JLzp}{J_{L,0}^p}

\newcommand{\JLnh}{J_{L,n}^h}

\newcommand{\JLnhpt}{J_{L,n+2}^h}
\newcommand{\JRnh}{J_{R,n}^h}
\newcommand{\JRnp}{J_{R,n}^p}
\newcommand{\NRnp}{N_{R,n}^p}

\newcommand{\NRnpp}{N_{R,n+1}^p}
\newcommand{\NRnpm}{N_{R,n-1}^p}
\newcommand{\NRnh}{N_{R,n}^h}
\newcommand{\NRnhp}{N_{R,n+1}^h}
\newcommand{\NRnhm}{N_{R,n-1}^h}
\newcommand{\NLnhp}{N_{L,n+1}^h}
\newcommand{\NLzp}{N_{L,0}^p}

\newcommand{\ALz}{A_{L,0}}

\newcommand{\NLnhpt}{N_{L,n+2}^h}

\newcommand{\ARn}{A_{R,n}}

\newcommand{\ALnpt}{A_{L,n+2}}

\newcommand{\tpn}{t_{p,n}}
\newcommand{\thn}{t_{h,n}}
\newcommand{\rhn}{r_{h,n}}
\newcommand{\rpn}{r_{p,n}}

\newcommand{\SLzp}{S_{L,0}^{p,\rightarrow}}

\newcommand{\SLnhl}{S_{L,n}^{h,\rightarrow}}
\newcommand{\SLnpl}{S_{L,n}^{p,\rightarrow}}
\newcommand{\SRnhr}{S_{R,n}^{h,\leftarrow}}
\newcommand{\SRnpr}{S_{R,n}^{p,\leftarrow}}
\newcommand{\SLnth}{S_{L,-2}^{h,\rightarrow}}
\newcommand{\Dpl}{D_{L}}
\newcommand{\Dl}{D_{L}}
\newcommand{\Dr}{D_{R}}

\newcommand{\Dpr}{D_{R}}

\newcommand{\Dpj}{D_{j}}

\newcommand{\qpj}{q_{p,j}}

\newcommand{\mulr}{\mu_{L/R}}
\newcommand{\mul}{\mu_{L}}
\newcommand{\mur}{\mu_{R}}
\newcommand{\Deltalr}{\Delta_{L/R}}
\newcommand{\Deltal}{\Delta_{L}}
\newcommand{\Deltar}{\Delta_{R}}
\newcommand{\nub}{\zeta}
\newcommand{\rarr}{\rightarrow}

\begin{document}

\title{Analytic approach to transport in superconducting junctions with arbitrary carrier density}

\author{F. Setiawan}
\email{setiawan@uchicago.edu}
\affiliation{Pritzker School of Molecular Engineering, University of Chicago, 5640 South Ellis Avenue, Chicago, Illinois 60637, USA}

\author{Johannes Hofmann}
\email{johannes.hofmann@physics.gu.se}
\affiliation{Department of Physics, Gothenburg University, 41296 Gothenburg, Sweden}

\begin{abstract}
Particle transport across junctions between two superconductors is commonly described using a simplifying approximation (often called the Andreev approximation), which assumes that excitations are fixed at the Fermi momentum and only Andreev reflections, with no normal reflections, occur at interfaces. While this approximation is appropriate for superconductors with high carrier density (for which the chemical potential vastly exceeds the pairing gap), it breaks down for superconductors with low carrier density, such as topological superconductors, doped semiconductors, or superfluid quantum gases. Here, we present a general \textit{analytical} framework for transport in superconducting junctions that does not rely on this limiting Andreev approximation. We apply our framework to describe transport in junctions between $s$-wave superconductors along the BCS-BEC crossover, which interpolates between the conventional high-carrier-density \mbox{(BCS-)regime} and moderate- as well as low-carrier-density regimes (unitary and BEC regimes), for which the high-carrier-density (Andreev) approximation is not valid. As the system is tuned from the BCS to the BEC regime, we find that the overall magnitude of a subgap current, which is attributed to multiple Andreev reflections, decreases. However, nonlinearities in the current-voltage characteristic become more pronounced near the intermediate unitary limit, giving rise to sharp peaks and dips in the differential conductance with even \textit{negative} differential conductance at specific voltages. Microscopically, the negative differential conductance is related to the van Hove points in the band structures, at which enhanced normal reflection occurs and that become accessible only when the chemical potential is comparable to or smaller than the pairing gap. The subgap current due to multiple Andreev reflections vanishes at a critical interaction strength on the BEC side, which we identify as the splitting point where the particle dispersion changes curvature. Our work shows that a description of transport in low-density superconducting junctions necessarily requires a treatment beyond the standard Andreev approximation.
\end{abstract}

\maketitle

\section{Introduction}

\begin{figure}[h!]
\includegraphics[width=\linewidth]{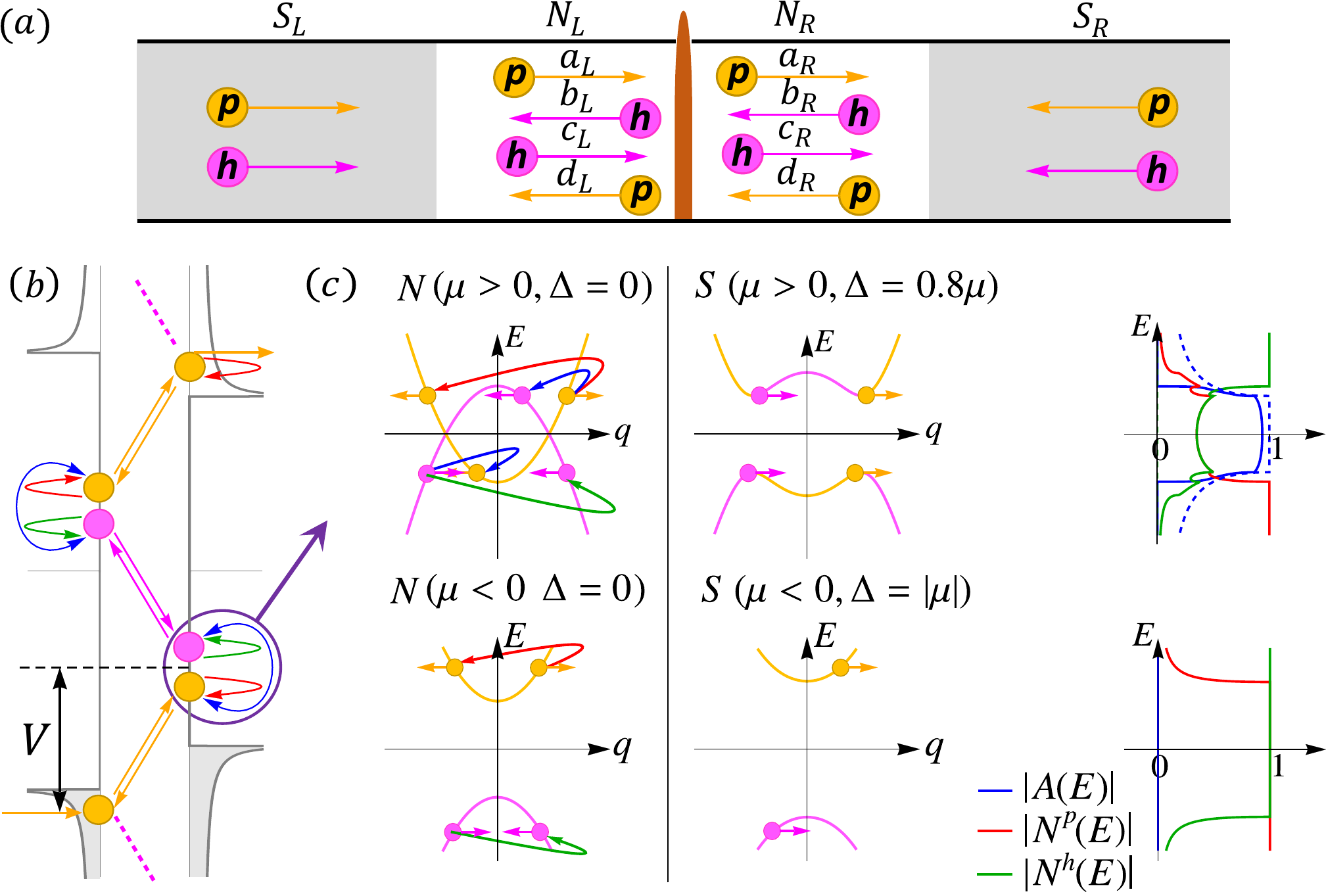}
\caption{(a) Superconducting junction with a tunnel barrier (brown vertical bar). Shown are the scattering amplitudes for particles ($p$, orange) and holes ($h$, magenta) with arrows indicating propagation directions. (b) A quasiparticle enters the junction from the left reservoir and gains energy through multiple Andreev reflections until it is transmitted into a reservoir. (c) Scattering at a normal-superconductor interface with \mbox{$\mu > 0$ and $\mu \simeq \Delta$} (upper panels) and \mbox{$\mu < 0$ with $|\mu| \simeq \Delta$} (lower panels). Left panels: Bogoliubov dispersions and examples of quasiparticle states. Right panels: Amplitudes of Andreev reflection [$A(E)$, blue line], normal reflection of particles [$N^p(E)$, red] and holes [$N^h(E)$, green]; Andreev approximation results are plotted as  dashed lines.}
\label{fig:1}
\end{figure}

Superconducting junctions, in which two superconductors are connected through a tunnel barrier or point contact [Fig.~\ref{fig:1}(a)], are the elementary building blocks of superconducting qubits~\cite{kjaergaard2020superconducting,Gyenis2021Moving} and quantum electronics~\cite{barone1982physics} with wide-ranging applications in high-precision measurements~\cite{popel1992josephson,braginski2004squid} and quantum computing~\cite{Alexeev2021Quantum}. Their properties are probed experimentally by measuring the current across the junction: For example, such measurements were among the first to characterize the superconducting gap~\cite{giaever60a,giaever60b}, and are now routinely used to determine the order parameter of exotic superconductors~\cite{deutscher05}, detect subgap states~\cite{Eichler2007Even,pillet2010andreev,Chang2013Tunneling,Ji2008High,Ruby2015Tunneling} such as Majorana bound states~\cite{Ruby2015End,feldman2017high}, or study semiconductor-superconductor heterostructures~\cite{gunel2012supercurrent,nilsson2012supercurrent,goffman2017conduction,gul2017hard,de2018spin,Kjaergaard2017Transparent}. 

%++++++++++++++++++++++++++++++++++++++++
%++++++++++++++++++++++++++++++++++++++++

Transport across superconducting junctions at bias voltages smaller than the tunneling gap occurs through a mechanism called Andreev reflection~\cite{degennes63,andreev64,saintjames64} where, at a normal-superconductor interface, a particle is reflected as a hole and vice versa. Extending transport calculations in normal-superconducting junctions~\cite{blonder82} to superconductor-normal-superconductor junctions  [Fig.~\ref{fig:1}(a)] is very difficult as one needs to account for multiple Andreev reflections~\cite{klapwijk82,octavio83} [see Fig.~\ref{fig:1}(b)], which occur repeatedly until a particle or hole acquires enough energy to overcome the excitation gap and transmits into either the left
or the right superconductor. In treating this process, existing theoretical calculations~\cite{averin95,hurd96,hurd97,arnold87,gunsenheimer94,cuevas96,bolech05} all rely on a high-carrier-density approximation (called the Andreev approximation)~\cite{andreev64,kulik70}, which assumes that the chemical potential of the superconductor is much larger than its pairing gap ($\mu \gg \Delta$), so that all particles have the same Fermi momentum and no normal reflections occur at normal-superconductor interfaces [red and green arrows in Fig.~\ref{fig:1}(b)]. This approximation  works extremely well for standard BCS-type superconductors like niobium~\cite{ludoph00}, but it is not reliable for most superconductors of current interest, such as topological superconductors~\cite{lutchyn10,oreg10,san2013multiple,Zazunov2016Low,setiawan17a,setiawan17b,heedt2021shadow}, high-$T_c$ superconductors~\cite{Golubov2000Andreev,deutscher05}, superconducting semiconductors like strontium titanate~\cite{richter13,cheng15,thierschmann2018transport,jouan2020quantized,briggeman2020pascal,mikheev20}, iron chalcogenides~\cite{kasahara14,kasahara16,rinott17,nakagawa21}, twisted cuprated superconductors~\cite{can2021high,Zhu2021Presence,zhao2021emergent} and superfluid junctions in ultracold Fermi gases~\cite{krinner15,valtolina15,husmann15,burchianti18,husmann18,Lebrat2018Band,Xhani2020Critical,kwon2020strongly,luick2020ideal,pace21}. The description of these systems requires a more general formulation of transport theory valid at all carrier densities. 

In this paper, we present a general analytical framework for transport in superconducting/superfluid junctions without resorting to a high-carrier-density approximation. Our calculations are based on the Landauer-B\"uttiker formalism, which describes the current across a Josephson junction in terms of Bogoliubov scattering states. One particular advantage of this approach is that it allows a discussion of the current in terms of elementary processes, such as the scattering of particles and holes at normal-superfluid interfaces and tunnel barriers, and thus provides an intuitive microscopic interpretation of transport features. To apply our approach to a system of current interest, we consider particle transport along the universal BCS-BEC crossover~\cite{zwerger11} in junctions between $s$-wave superconductors or superfluids, which describes transport in fermionic quantum gases and superconducting semiconductors like strontium titanate. Here, the reservoir interaction strength is parametrized by an inverse $s$-wave scattering length $(k_F a)^{-1}$, where $k_F$ is the Fermi momentum, which continuously interpolates between a high carrier-density BCS limit \mbox{$(k_F a)^{-1} \ll 0$} with \mbox{$\mu \gg \Delta$} and regimes where existing calculations break down: a unitary regime at \mbox{$(k_F a)^{-1} = 0$} with \mbox{$\mu \simeq \Delta$} followed by a BEC regime at \mbox{$(k_F a)^{-1} \gg 0$} with negative chemical potential $\mu < 0$.
 
Our general formalism reveals two features not captured within the high-carrier-density approximation: First, we show that multiple Andreev reflections lead to a significantly more pronounced current-voltage characteristic in the unitary regime compared to the BCS limit. Here, the current shows sharp peaks and dips at specific voltages that lead to \textit{negative} differential conductance, which are associated with the van Hove points of the band structures at which enhanced normal reflections occur. Second, the subgap current induced by multiple Andreev reflections vanishes at a critical interaction strength on the BEC side of the crossover, and we identify this point as the so-called splitting point~\cite{son06,haussmann09,frank18} where the quasiparticle dispersion changes its curvature (i.e., its minimum shifts from finite to zero momentum). 

This paper is structured as follows: In Sec.~\ref{eq:framework}, we outline the general Landauer-B\"uttiker framework that applies beyond the high carrier-density (Andreev) approximation. We begin in Sec.~\ref{eq:setup} by presenting a general Josephson junction setup and introducing basic definitions. In Sec.~\ref{sec:landauer}, we derive an expression for the current across the junction. Section~\ref{sec:scatteringmatrix} presents the scattering states used to calculate the current. In particular, we derive a closed-form recurrence relation linking the scattering coefficients at different energies, which are connected by multiple Andreev reflection processes. The solution of the recurrence relation is sketched in Sec.~\ref{sec:numerics}. As an important check of the validity of our solution, the tunneling limit is discussed in Sec.~\ref{sec:tunneling}. Section~\ref{sec:results} presents results for the dc current and differential conductance in a Josephson junction across the BCS-BEC crossover; readers interested only in the applications of our method can skip directly to this section. In Secs.~\ref{sec:negativeconductance} and~\ref{sec:splitting}, we discuss in detail our main results, which cannot be obtained using the Andreev approximation, namely the negative differential conductance at unitarity and the suppression of the multiple-Andreev-reflection induced subgap current as the interaction is tuned to the BEC regime. We end with a conclusion in Sec.~\ref{sec:conclusions}. Various details of the calculation and analytical results are relegated to Appendices~\ref{sec:bogoliubov} to~\ref{sec:SGS}.

\section{Theoretical Framework}\label{eq:framework}

This section presents a self-contained derivation of the Landauer-B\"uttiker formalism for superfluid-normal-superfluid junctions, where we focus in particular on complications introduced when going beyond the Andreev approximation. General introductions to the Landauer-B\"uttiker formalism applied to superconducting junctions are found, for example, in Refs.~\cite{datta96,lesovik11}. Earlier references that discuss transport across superfluid-normal-superfluid junctions using the Andreev approximation are Refs.~\cite{bratus95,averin95,hurd96,hurd97}.

\subsection{Setup}\label{eq:setup}

We  consider two superfluid reservoirs as sketched in Fig.~\ref{fig:1}(a). The reservoir on the left-hand side (right-hand side) is defined by a chemical potential $\mu_L$ ($\mu_R$) and a pairing gap $\Delta_L$ ($\Delta_R e^{i \phi}$). There is a potential $U(x, {\bf r}_\perp)$ that confines particles in the radial ${\bf r}_\perp$ direction but allows particle exchange between the reservoirs through a small contact at $x = 0$, thereby giving rise to a particle current. Given a bias $V = \mu_L - \mu_R$, we aim to determine this current, i.e., the expectation value of the operator
\begin{align}
\hat{I}_x(\tau, x) &= \frac{-i \hbar}{2m} \sum_{\sigma=\uparrow,\downarrow} \int d{\bf r}_\perp \Bigl[\hat{\psi}_\sigma^\dagger (\nabla \hat{\psi}_\sigma^{}) - (\nabla \hat{\psi}_\sigma^\dagger) \hat{\psi}_\sigma^{}\Bigr] , \label{eq:particlecurrent}
\end{align}
where $\hat{\psi}_\sigma$ is a fermion field operator for a particle of mass $m$ and spin projection $\sigma$. The Landauer-B\"uttiker formalism assumes no particle interactions inside the junction (for a study of interaction effects in a quantum dot in the BCS-BEC crossover using other methods, see Ref.~\cite{hofmann17}) and describes transport in terms of scattering solutions for Bogoliubov excitations. In principle, the condensate depends self-consistently on the Bogoliubov modes through a gap equation,
\begin{align}
\Delta({\bf r}) &= - \frac{g}{2} \sum_n f(E_n) \Phi_n^\dagger({\bf r}) \sigma_x \Phi_n({\bf r}) , \label{eq:Deltagap}
\end{align}
where $g$ is the interaction strength, $f(E)= 1/[\mathrm{exp}(E/k_{\mathrm{B}}T)+1]$ is the Fermi function (we set the temperature $T = 0$), $n$ runs over all Bogoliubov eigenstates with excitation energy $E_n$, $\sigma_x$ is the $x$-Pauli matrix in Nambu space, and $\Phi_n$ solves the Bogoliubov-de Gennes equation
\begin{align}\label{eq:bdg}
E_n\Phi_n({\bf r}) &= \begin{pmatrix}H_0 - \mu({\bf r}) & \Delta({\bf r}) \\ \Delta^*({\bf r}) & - (H_0^* - \mu({\bf r})) \end{pmatrix} \Phi_n({\bf r}) . 
\end{align}
Here, $H_0 = -\hbar^2 \nabla^2/2m + U(x, {\bf r}_\perp)$ is the Hamiltonian of a single particle in the potential $U(x, {\bf r}_\perp)$. If the confining potential $U(x,\mathbf{r}_\perp)$ varies slowly along the $x$ direction, we may separate the wave function into a transverse part $\lambda_\alpha({\bf r}_\perp)$ and a longitudinal part $\Psi_{n}(x)$ as $\Phi_n({\bf r}) = \Psi_{n}(x) \lambda_\alpha({\bf r}_\perp)$, where $\alpha$ is the transverse band index. In the following, we consider a single transverse band. 

In the standard description of Josephson junctions, we assume that the chemical potential is given by a step-like function
\begin{align}
\mu(x) &= \begin{cases}\mul, & x <0, \\[1ex]  \mur, & x > 0. \end{cases} \label{eq:mupotential}
\end{align}
In addition, we include a tunnel barrier at $x=0$ to account for the transparency of the junction. 
In a self-consistent calculation, the gap inside the constriction will be smaller than the bulk value in the respective reservoirs~\cite{spuntarelli10} and it will be further decreased by beyond mean-field fluctuations in this confined geometry~\cite{larkin05}. 
In this paper, we thus make the standard approximation~\cite{datta96,lesovik11} to forgo a self-consistent solution of Eq.~\eqref{eq:Deltagap} and instead choose a step-like function for the pairing gap
\begin{align}
\Delta(x) &= \begin{cases}\Deltal, & x\leq -\ell/2, \\ 0, & -\ell/2<x <\ell/2, \\ \Deltar e^{i\phi}, & x \geq \ell/2, \end{cases} \label{eq:potential}
\end{align}
with constant pairing gaps in both reservoirs (in the following, we set the superconducting phase difference $\phi =0$) equal to their bulk values, while the constriction is assumed to be a normal region of length $\ell$. Since in a non-self-consistent description of the pairing potential, the current in the superconducting region is not a conserved quantity~\cite{Datta1996scattering}, the inclusion of this normal region simplifies the calculation as the current can be evaluated in the normal region, where current conservation holds. The step-function profile~\eqref{eq:potential} has the additional advantage that it allows for a transparent discussion of elementary physical processes, i.e., Andreev and normal reflections at the normal-superfluid interfaces as well as scattering off the tunnel barrier in the normal region, where the scattering matrices corresponding to these processes can be derived analytically. The step-profile~\eqref{eq:potential} is a conventional approximation for superconducting junctions that is used widely in the literature~\cite{averin95,hurd96,hurd97,arnold87,gunsenheimer94,cuevas96,bolech05,kulik70,ludoph00,san2013multiple,Zazunov2016Low,setiawan17a,setiawan17b}, and it is valid provided that the junction length $\ell$ is shorter than the coherence length (or specifically the phase coherence length) of the superconductors since details of the junction do not matter in this regime. For simplicity, we consider ``short junctions" ($\ell \rightarrow 0$) throughout the paper.

\subsection{Landauer-B\"uttiker formalism}\label{sec:landauer}

The Landauer-B\"uttiker formalism makes two central assumptions:

(a) The fermion operators in Eq.~\eqref{eq:particlecurrent} may be expanded in a basis set of Lippmann-Schwinger scattering states across the junction [Eqs.~\eqref{eq:mupotential} and~\eqref{eq:potential}]. Restricting to a single transverse channel $\alpha=0$ with transverse eigenmode $\lambda_0({\bf r}_\perp)$, these fermion operators are
\begin{align}
&\begin{pmatrix}
\hat{\psi}_\uparrow^{}(\tau, {\bf r}) \\[0.5ex]
\hat{\psi}_\downarrow^\dagger(\tau, {\bf r})
\end{pmatrix}\nonumber\\
&\quad= \lambda_0({\bf r}_\perp) 
\sum_{\substack{\kappa=p,h \\ j=L,R}} 
\int_{-\infty}^\infty \frac{dE}{\sqrt{2\pi \vFermikj(E)}} \, \sqrt{\frac{\hbar q_{\kappa,j}(E)}{m}}  \nonumber \\
&\qquad\times
\bigg(
\Psikj^{S,\rightarrow}(E, \tau, x) \akj^\rightarrow(E) +
\Psikj^{S,\leftarrow}(E, \tau, x) \akj^\leftarrow(E)
\biggr) , \label{eq:lippmannschwinger}
\end{align}
where $\Psikj^{S,\zeta}(E, \tau, x)$ describes the time-dependent scattering states in the longitudinal direction for the reservoir \mbox{$j=L,R$} that are asymptotically described by a right-moving ($\zeta = \rightarrow$) or left-moving ($\zeta=\leftarrow$) incoming Bogoliubov quasiparticles ($\kappa = p$) or quasiholes ($\kappa = h$) with energy $E$. We postpone the explicit calculation of these scattering states to the next section (Sec.~\ref{sec:scatteringmatrix}). The fermionic Bogoliubov operators in Eq.~\eqref{eq:lippmannschwinger} satisfy the anticommutation relation
\begin{align}
\{\akj^\zeta(E), \hat{a}_{\kappa',j'}^{\zeta'}(E')\} =\delta_{\kappa\kappa'} \delta_{jj'}  \delta_{\zeta \zeta'} \delta(E-E').
\end{align}
Furthermore, the term
\begin{align}
\vFermikj(E) &= \biggl| \frac{d\qkj(E)}{dE} \biggr|^{-1}
\end{align}
in Eq.~\eqref{eq:lippmannschwinger} is the residual Jacobian of the transformation from a wave number to energy integration. By convention, we separate a factor $\sqrt{\hbar \qkj(E)/m}$ in Eq.~\eqref{eq:lippmannschwinger} from the Lippmann-Schwinger wave functions such that the Bogoliubov excitations carry unit probability current.

(b) It is assumed that the scattering states are in equilibrium with their respective reservoirs, i.e., 
\begin{align}
\langle \akj^{\zeta\dagger}(E) \hat{a}_{\kappa',j'}^{\zeta'}(E') \rangle &=  \delta_{\kappa\kappa'}\delta_{jj'} \delta_{\zeta\zeta'} \delta(E-E') f(E) ,
\end{align}
where $f(E)$ is the Fermi-Dirac distribution. 

The expectation value of the current [Eq.~\eqref{eq:particlecurrent}] consists of a contribution from quasiparticles and quasiholes injected from the left reservoir \mbox{$j=L$} (calculated using the scattering states $\Psi_{\kappa,L}^{S,\rightarrow}$) and a current due to quasiparticles and quasiholes injected from the right reservoir \mbox{$j=R$} (calculated using the state $\Psi_{\kappa,R}^{S,\leftarrow}$). 
The total current is the difference
\begin{equation}
I(V) =  I^{\rightarrow}(V) - I^{\leftarrow}(V) \label{eq:currentdiff}
\end{equation}
between a current $I^{\rightarrow}$ from the left to the right reservoir and its reverse $I^{\leftarrow}$. 
Since states injected from the left/right reservoir that impinge on the junction are right/left moving, 
we will use \mbox{$j=L,R$} and suppress the index~$\leftrightharpoons$ in some of the following discussions for notational simplicity. Because of the chemical potential mismatch between the left and right reservoirs, states at different energies are related by multiple Andreev reflections in the Josephson junction. 
The Lippmann-Schwinger scattering states are then given by
\begin{align}\label{eq:Psikj}
\Psikj^S(E, \tau,x) &= \sum_{n=-\infty}^\infty e^{-iE_n\tau/\hbar} 
\begin{pmatrix}
\varphi_{\kappa,j}^{(n)}(x)
\\[1ex]
\chi_{\kappa,j}^{(n)}(x)
\end{pmatrix} , 
\end{align}
where 
\begin{align}
E_n = E + n V \label{eq:En},
\end{align}
with $V=\mu_L-\mu_R$ being the chemical potential difference between the two reservoirs, i.e., the bias voltage. Substituting Eq.~\eqref{eq:Psikj} into Eq.~\eqref{eq:particlecurrent}, we write the current as 

\begin{align}
I &= I_{\rm dc}(V) + 2\sum_{l=2,4,6,\cdots} \bigl[\mathcal{A}_l(V) \cos (\phi_l + \omega_l \tau) \nonumber \\
&\qquad\qquad\qquad\qquad\qquad+ \mathcal{B}_l(V) \sin (\phi_l+\omega_l \tau) \bigr] ,
\end{align}
where $\phi_l = l\phi/2$, $\omega_l \equiv lV/\hbar$, and
\begin{widetext}
\begin{subequations}
\label{eq:currentfourier}
\begin{align}
\mathcal{A}_l(V) &= \frac{2}{h} {\rm Re} \, \int_{-\infty}^\infty dE  \biggl\{\Dpl(E) \, \bigl[f(E) T_{p\rightarrow}^{p(l)}(E)+ (1-f(E)) T_{p\rightarrow}^{h(l)}(E)\bigl]-\Dpr(E) \bigl[f(E) T_{p\leftarrow}^{p(l)}(E)+ (1-f(E)) T_{p\leftarrow}^{h(l)}(E)\bigl]\biggl\},\\
\mathcal{B}_l(V) &= \frac{2}{h} {\rm Im} \, \int_{-\infty}^\infty dE   \biggl\{ \Dpl(E) \, \bigl[f(E) T_{p\rightarrow}^{p(l)}(E)+ (1-f(E)) T_{p\rightarrow}^{h(l)}(E)\bigl]-\Dpr(E)  \bigl[f(E) T_{p\leftarrow}^{p(l)}(E)+ (1-f(E)) T_{p\leftarrow}^{h(l)}(E)\bigl]\biggl\}, \\
I_{\rm dc}(V) &= \mathcal{A}_0(V).
\end{align}
\end{subequations}
\end{widetext}
Here, 
\begin{align}\label{eq:Dpj}
\Dpj(E)&= \frac{\hbar^2}{m} \frac{\qpj(E)}{\vFermipj(E)} .
\end{align}
is an effective quasiparticle density of states in the $j =L, R$ reservoir, $T_{p,\zeta}^{p(l)}(E)$ and $T_{p,\zeta}^{h(l)}(E)$ are the dimensionless \mbox{$l$}th Fourier components of the particle and hole current density due to quasiparticle injections  at energy $E$ from the left ($\zeta = \rightarrow$) or the right ($\zeta =\leftarrow$) superfluids, which are given by
\begin{subequations}\label{eq:Tphm}
\begin{align}
T_{p,\rightarrow/\leftarrow}^{p(l)}(E) &= 
\frac{- i \hbar}{2 m} \sum_{n=-\infty}^\infty \Bigl(
\varphi_{p,L/R}^{(n+l)*}(x) [\nabla \varphi_{p,L/R}^{(n)}(x)] \nonumber \\
& \qquad\qquad- [\nabla \varphi_{p,L/R}^{(n+l)*}(x)] \varphi_{p,L/R}^{(n)}(x)
\Bigr), \\
T_{p,\rightarrow/\leftarrow}^{h(l)}(E) &= 
\frac{- i \hbar}{2 m} \sum_{n=-\infty}^\infty \Bigl(
\chi_{p,L/R}^{(n)}(x) [\nabla \chi_{p,L/R}^{(n+l)*}(x)] \nonumber \\
& - [\nabla \chi_{p,L/R}^{(n)}(x)] \chi_{p,L/R}^{(n+l)*}(x)
\Bigr).
\end{align}
\end{subequations}
As shown in Sec.~\ref{sec:scatteringmatrix}, the quasihole contribution to the current is equal to the quasiparticle current, hence the overall factor of $2$ in Eq.~\eqref{eq:currentfourier}. 
While in this paper, we focus only on the dc current, our formalism can also be used to calculate the time-dependent (ac) current by computing the coefficients $\mathcal{A}_l(V)$ and $\mathcal{B}_l(V)$ in Eq.~\eqref{eq:currentfourier}, where $l = 2,4,6,\ldots$ is a positive even integer. 

\subsection{Scattering matrix formalism}\label{sec:scatteringmatrix}

To evaluate the current densities [Eq.~\eqref{eq:Tphm}], we need to calculate the scattering states in Eq.~\eqref{eq:lippmannschwinger} by solving the Bogoliubov equation [Eq.~\eqref{eq:bdg}] with spatially varying potentials $\mu(x)$ and $\Delta(x)$ given in Eqs.~\eqref{eq:mupotential} and~\eqref{eq:potential}. In this section, we derive these states and obtain an expression for the current density.

Within each region, where $\mu$ and $\Delta$ are constant, solutions take a standard plane-wave form with Bogoliubov dispersion 
\begin{align}\label{eq:energyq}
E^2 = (\varepsilon_q-\mu)^2 + |\Delta|^2 ,
\end{align}
where \mbox{$\varepsilon_{q} = \hbar^2 q^2/2m$}. Figure~\ref{fig:1}(c) illustrates this dispersion in the superfluid and normal regions for \mbox{$\mu= 1.25\Delta$} (upper panel) and \mbox{$\mu=-\Delta$} (lower panel), where arrows indicate the propagation direction set by the group velocity $\tilde{v}(E) = \partial E(q)/\partial q$. Here, we distinguish particle-type (orange) and hole-type (magenta) excitations, for which the velocity and the momentum have equal or opposite sign, respectively. Within a high carrier-density approximation ($\mu \gg \Delta$), this picture simplifies considerably: All excitations are fixed at the Fermi momentum \mbox{$ \hbar k_F = \sqrt{2m\varepsilon_F}$}, and the momentum essentially drops out as a variable. 

%++++++++++++++++++++++++++++++++++++++++
\begin{figure}[t]
\includegraphics[width=\linewidth]{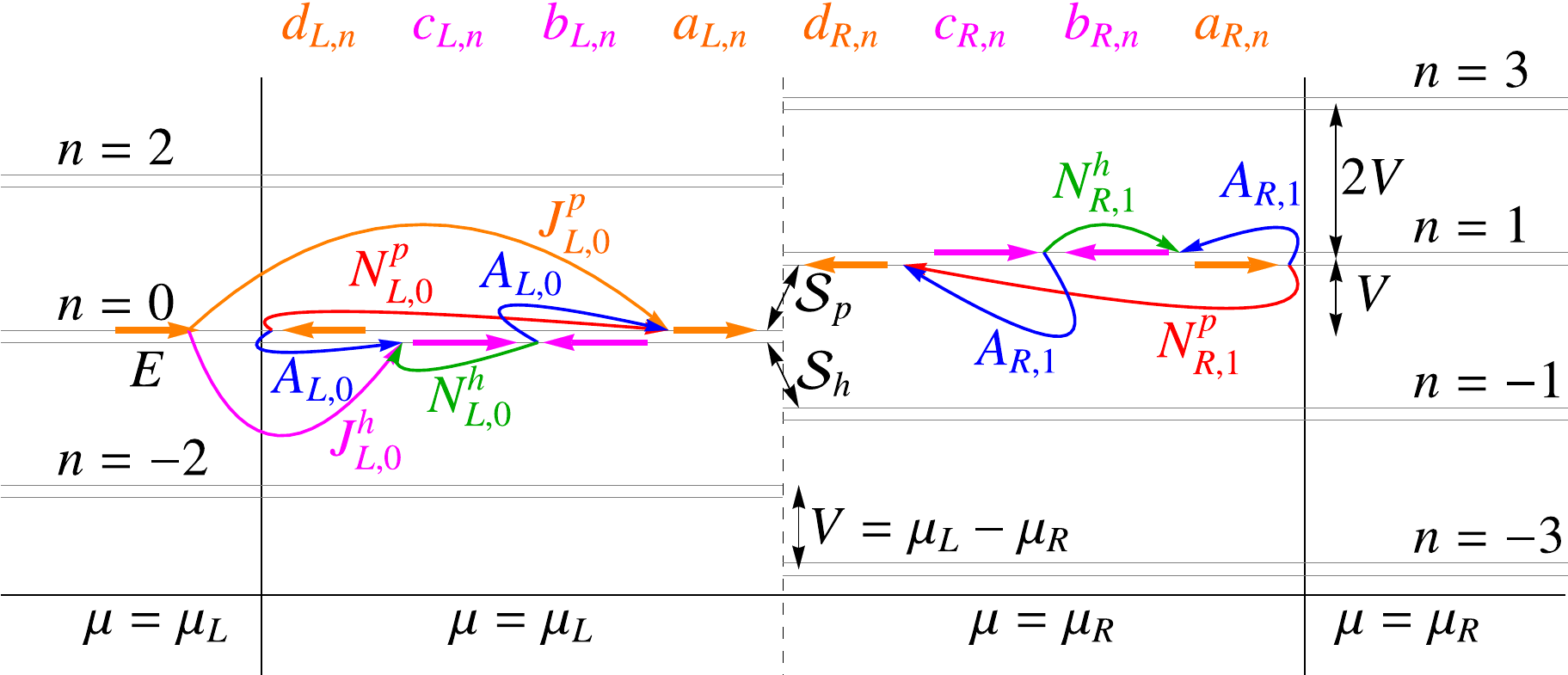}
\caption{
Scattering processes across the superfluid-normal-superfluid junction, where solid-vertical lines mark the normal-superfluid boundaries and the dashed line denotes the tunnel barrier. Shown is the insertion of particle ($J^p_{L,n}$) and hole ($J^h_{L,n}$) states into the left normal region due to the injection of a quasiparticle from the left superfluid reservoir with energy $E$ as well as subsequent scattering processes. Normal reflections of particles ($N_{L/R,n}^{p}$) and holes ($N_{L/R,n}^{h}$) at the normal-superfluid boundary are indicated by red and green arrows, respectively, while Andreev ($A_{L/R,n}$) reflections are denoted by blue arrows [the same convention is used as in Fig.~\ref{fig:1}]. The magenta- and orange-colored horizontal arrows correspond to the scattering states for particles and holes, respectively, where in our convention the voltage drop across the reservoirs is absorbed in a time-dependent tunnel barrier that changes the energy of particles and holes. At the tunnel barrier, each transmission event changes the Floquet index $n$ of scattering states by $\pm 1$ where the scattering matrices at the tunnel barrier are denoted by $\mathcal{S}_p$ for particles and $\mathcal{S}_h$ for holes.
}
\label{fig:scatteringstates}
\end{figure}
%++++++++++++++++++++++++++++++++++++++++

The particle/hole-like (\mbox{$\kappa = p/h$}) Bogoliubov state with energy $E$ (normalized to unit probability current) in the left ($L$) and right ($R$) reservoir is
\begin{eqnarray}
\Psi^S_{\kappa L}(E) &=& e^{- i \frac{E t}{\hbar}} \sqrt{\frac{m}{\hbar q_{\kappa L}(E)}} 
\biggl(\begin{matrix} u_{\kappa L}(E) \\v_{\kappa L}(E) \end{matrix}\biggr)
e^{i q_{\kappa L} x} , \label{eq:bogoliubovL} \\
\Psi^S_{\kappa R}(E) &=& e^{- i \frac{E t}{\hbar}} \sqrt{\frac{m}{\hbar q_{\kappa R}(E)}} 
\biggl(\begin{matrix} u_{\kappa R}(E) e^{- i\frac{V t}{\hbar}} \\v_{\kappa R}(E) e^{+ i \frac{V t}{\hbar}}\end{matrix}\biggr) e^{i q_{\kappa R}x} , \quad \label{eq:bogoliubovR}
\end{eqnarray}
with Bogoliubov factors $u_{pL/R}^2, v_{pL/R}^2 = [1 \pm (\varepsilon_{q} - \mu_{L/R})/E]/2$. Detailed properties of the Bogoliubov states are reviewed in Appendix~\ref{sec:bogoliubov}. The time dependence $e^{\mp i V t/\hbar}$ in Eq.~\eqref{eq:bogoliubovR} appears because the state is written with reference to the left reservoir chemical potential~\cite{datta96}. As discussed in the previous section, due to multiple Andreev reflections the scattering state is a superposition of  right- and left-moving Bogoliubov states with energy $E_n$ given in Eqs.~\eqref{eq:En} and~\eqref{eq:energyq},  
where the amplitudes must be matched at the normal-superfluid boundaries and tunnel barrier.   We use $a_{L/R,n}$, $b_{L/R,n}$, $c_{L/R,n}$, and $d_{L/R,n}$ to denote the amplitudes of right-moving particles, left-moving holes, right-moving holes and left-moving particles with energy $E_n$ in the left ($L$)/right ($R$) normal region, respectively. This is sketched in Fig.~\ref{fig:scatteringstates}, which shows the energy-resolved structure of different contributions to the scattering wave function for a state inserted from the left reservoir. As is apparent from the figure, computing the Lippmann-Schwinger states in the presence of a voltage bias is a highly nontrivial Floquet-type problem rather than a simple potential scattering calculation. The full form of the scattering wave functions is listed in Appendix~\ref{sec:scatteringstates}.

Substituting the scattering wave functions, we evaluate the dc current in Eqs.~\eqref{eq:currentfourier} and \eqref{eq:Tphm} as
\begin{eqnarray}\label{eq:Idc}
I_{\rm dc}^{\rightarrow}(V) &=& \dfrac{2}{h} \int_{-\infty}^\infty dE \Dpl(E)\nonumber\\ 
&\times&\bigl[f(E) T_{p,\rightarrow}^{p(0)}(E)+ (1-f(E)) T_{p,\rightarrow}^{h(0)}(E)\bigl], \quad 
\end{eqnarray}
where $\Dpl(E)$ is the quasiparticle density of states of the left superfluid, and \\[-3ex]
\begin{subequations}\label{eq:Tpzeta}
\begin{eqnarray}
T_{p,\rightarrow}^{p(0)}(E) &=& {\displaystyle \sum_{n=-\infty}^{\infty}} (|a^{}_{L,n}|^2 - |d^{}_{L,n}|^2) \Theta(\mu_L + E_n),\qquad \label{eq:Tpzetaa}\\
T_{p,\rightarrow}^{h(0)}(E) &=& {\displaystyle \sum_{n=-\infty}^{\infty}} (|b^{}_{L,n}|^2 - |c^{}_{L,n}|^2) \Theta(\mu_L - E_n)  \label{eq:Tpzetab}
\end{eqnarray}
\end{subequations}
are the dimensionless particle and hole current densities at energy $E$ due to quasiparticles that are transmitted into the normal region as particles [Eq.~\eqref{eq:Tpzetaa}] and holes [Eq.~\eqref{eq:Tpzetab}], respectively. Due to particle-hole symmetry, there is an equal current from quasihole injections, hence the factor of $2$ in Eq.~\eqref{eq:Idc}. 

%++++++++++++++++++++++++++++++++++++++++
\begin{figure*}[t]
\includegraphics[width=\linewidth]{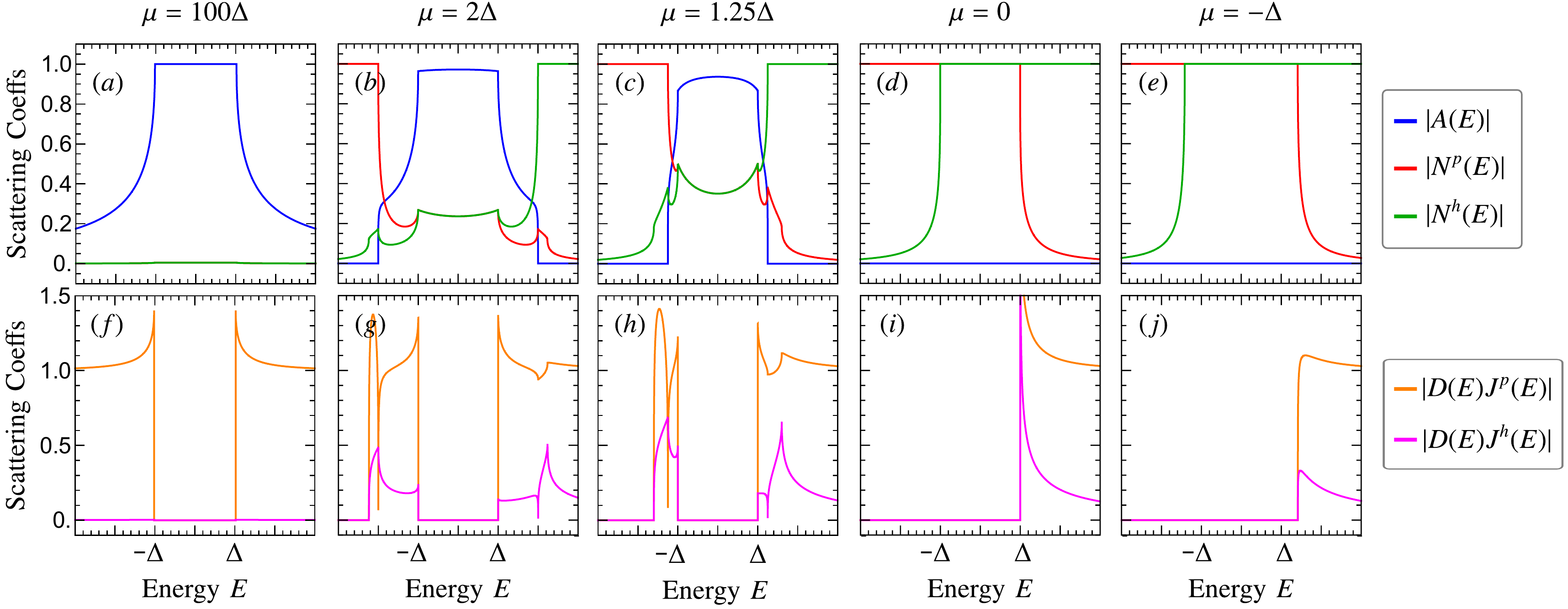}
\caption{Plots of dimensionless scattering coefficients at a normal-superfluid interface for different ratios of chemical potential $\mu$ and gap $\Delta$. [Upper panel: (a)--(e)] Amplitudes for Andreev reflection [$A(E)$], normal reflections of particles [$N^p(E)$], and holes [$N^h(E)$]. Note that the magnitude of the normal reflection coefficients $N^p(E)$ and $N^h(E)$ for energies below the gap ($|E| < \Delta$) increases from 0 (no normal reflection) to 1 (total normal reflection) as the chemical potential changes from large positive values ($\mu \gg \Delta$) to negative values, i.e., as we move away from the Andreev approximation regime. [Lower panel: (f)--(j)]  Transmission amplitudes of quasiparticle injections from the reservoir into particles [$J^p(E)$] and holes [$J^h(E)$] in the normal region, multiplied by the quasiparticle density of states $D(E)$.  Note that $J^h(E)$ becomes nonzero as we move away from the Andreev approximation regime.}\label{fig:S4}
\end{figure*}
%++++++++++++++++++++++++++++++++++++++++
 
 There are several relations between the amplitudes of right-moving particle and hole states ($a_{n}$ and $c_{n}$) and left-moving states ($d_{n}$ and $b_{n}$) due to elementary scattering processes at the superfluid-normal interfaces or the tunnel barrier:

First, at the left S$_L$-N$_L$ boundary, particles or holes in the normal region that propagate away from the normal-superfluid boundary are created either by transmitting a Bogoliubov excitation across the boundary, normal 
-reflecting a particle or hole that impinges on the normal-superfluid junction, or by Andreev-reflecting an impinging excitation to an excitation of opposite type. This is described by the scattering matrix equation
\begin{equation}
\begin{pmatrix} a_{L,n}^{} \\ c_{L,n}^{} \end{pmatrix}
=
\begin{pmatrix} 
\NLnp & \ALn \\
\ALn & \NLnh
\end{pmatrix}
\begin{pmatrix} d_{L,n}^{} \\ b_{L,n}^{} \end{pmatrix} 
+ \delta_{n0} 
\begin{pmatrix} \JLp \\ \JLh\end{pmatrix} , \label{eq:resmatchingL}
\end{equation}
where $A_{L,n}$, $N_{L,n}^{p}$, and $N_{L,n}^{h}$ are the (energy-dependent) Andreev reflection as well as the normal reflection amplitude for particles and holes, respectively. Note that normal reflections can occur due to the breakdown of the Andreev approximation (as discussed in this paper), but can also arise due to a mismatch in Fermi velocities~\cite{Mortensen1999Angle} or for $d$-wave pairing~\cite{Golubov2000Andreev,Bruder1990Andreev}. Here, $\JLp$ and $\JLh$ are the transmission amplitudes of injected quasiparticles from the left (S$_L$) reservoir into particles and holes in the left normal region (N$_L$).  Analytical expressions for these amplitudes are obtained by matching the wave functions at the normal-superfluid boundaries. Explicit expressions are listed in Appendix~\ref{sec:matching}. Note that in Eq.~\eqref{eq:resmatchingL}, we consider quasiparticle states injected from the left reservoir, which contribute to $I_{\rm dc}^{\rightarrow}$ (the reverse current $I_{\rm dc}^{\leftarrow}$ is obtained by interchanging the reservoir indices).  
 
 Second, the analogous process at the right N$_R$-S$_R$ boundary is
\begin{equation}
\begin{pmatrix} d_{R,n}^{} \\ b_{R,n}^{}\end{pmatrix}
=
\begin{pmatrix} 
\NRnp & \ARn \\
\ARn & \NRnh
\end{pmatrix}
\begin{pmatrix} a_{R,n}^{} \\ c_{R,n}^{} \end{pmatrix}. 
\label{eq:resmatchingR} 
\end{equation}
The scattering processes [Eqs.~\eqref{eq:resmatchingL} and~\eqref{eq:resmatchingR}] are illustrated in Figs.~\ref{fig:1}(b) and~\ref{fig:scatteringstates}, where Andreev reflections are sketched in blue, normal reflections of particles in red and of holes in green.  The expression for the coefficients $\NRnp$, $\NRnh$, $\ARn$, $\JRnp$, $\JRnh$ are obtained from the corresponding coefficients at the left S-N boundary by replacing the subscript $L$ by $R$. 
The corresponding scattering amplitudes together with the band structure at the junction interfaces are shown for two cases in Fig.~\ref{fig:1}(c). 
Within the Andreev approximation, a quasiparticle entering from a reservoir is always transmitted into the normal region as a particle, such that $J_L^p = 1$ and $J_L^h = 0$; the opposite is true for a quasihole. In addition, normal reflection is absent and perfect Andreev reflection $|A_{L,n}| = 1$ occurs for energies below the pairing gap [blue dashed line in Fig.~\ref{fig:1}(c)]. As is apparent from the figure, the Andreev approximation is not reliable once \mbox{$\mu \simeq \Delta$}. 

Figure~\ref{fig:S4} shows in more detail the scattering coefficients for different ratios of the chemical potential $\mu$ and gap $\Delta$, where results in the Andreev approximation regime are shown in Figs.~\ref{fig:S4}(a) and~\ref{fig:S4}(f). Note that Figs~\ref{fig:S4}(c) and~\ref{fig:S4}(e) are identical to the right upper and lower panels in Fig.~\ref{fig:1}(c). As is apparent from the analytical results [Eqs.~\eqref{eq:scattcoeff} and~\eqref{eq:JLn} in Appendix~\ref{sec:matching}] and Fig.~\ref{fig:S4}, the expressions simplify considerably in the Andreev approximation limit since there is no normal reflection for particles/holes, and Bogoliubov quasiparticles inserted into the normal region are always transmitted as particles (never as holes) and vice versa. Furthermore, we note that the Andreev approximation becomes unreliable even for a moderate increase in the gap [see Figs.~\ref{fig:S4}(b) and~\ref{fig:S4}(g)] and no longer gives the correct scattering form as we deviate further from the Andreev approximation regime [see Figs.~\ref{fig:S4}(c)--\ref{fig:S4}(e) and~\ref{fig:S4}(h)--\ref{fig:S4}(j)].

Third, the scattering matrices for particles and holes at the tunnel barrier are
\begin{subequations}\label{eq:scatbarrier}
\begin{align}
\begin{pmatrix} d_{L,n}^{} \\ a_{R,n+1}^{} \end{pmatrix}
&= \mathcal{S}_p \begin{pmatrix} a_{L,n}^{} \\ d_{R,n+1}^{} \end{pmatrix}= \begin{pmatrix}r_{p,n} & t_{p,n} \\ t_{p,n} & - \frac{t_{p,n}}{t_{p,n}^*} r_{p,n}^*\end{pmatrix} \begin{pmatrix} a_{L,n}^{} \\ d_{R,n+1}^{} \end{pmatrix},\label{eq:particlebarrier}\\ 
\begin{pmatrix} b_{L,n}^{} \\ c_{R,n-1}^{} \end{pmatrix}
&= \mathcal{S}_h \begin{pmatrix} c_{L,n}^{} \\ b_{R,n-1}^{} \end{pmatrix}= \begin{pmatrix}r_{h,n} & t_{h,n} \\ t_{h,n} & - \frac{t_{h,n}}{t_{h,n}^*} r_{h,n}^*\end{pmatrix} \begin{pmatrix} c_{L,n}^{} \\ b_{R,n-1}^{} \end{pmatrix} \label{eq:holebarrier},
\end{align}
\end{subequations}
with transmission ($\tpn$, $\thn$) and reflection coefficients ($\rpn$, $\rhn$). In this paper, we assume a delta-function barrier~\cite{griffiths18,lesovik11} with energy-independent reflection and transmission coefficients given by
\begin{align}
t_{p,n} = t_{h,n}^* = te^{i\eta} \label{eq:tpn}
\end{align}
and 
\begin{align}
r_{p,n} = r_{h,n}^*  = -ie^{i\eta}\sqrt{1-t^2} \label{eq:rpn}
\end{align}
where $\eta = -\mathrm{arctan}(Z)$. Here, $Z$ is the dimensionless barrier strength as defined in the Blonder-Tinkham-Klapwijk theory~\cite{blonder82} which is related to the barrier transparency $\mathcal{T} = t^2$ by $Z = \sqrt{(1/\mathcal{T})-1}$.

The various coefficients in the scattering wave functions are successively eliminated using Eqs.~\eqref{eq:resmatchingL}-\eqref{eq:scatbarrier}, which reduce to a recurrence relation for a single set of coefficients $\{d_{L,n}\}$ in the normal region. A lengthy but straightforward calculation, which is detailed in Appendix~\ref{sec:recur}, gives the following recurrence relation:
\begin{equation}
\alpha_n^{} d_{L,n+ 2}^{} + \beta_n^{} d_{L,n}^{} + \gamma_n^{} d_{L,n - 2}^{} = S_L^p  \delta_{n0} + S_L^h \delta_{n,-2} , \label{eq:recurrence} 
\end{equation}
where $\alpha_n^{}$, $\beta_n^{}$, $\gamma_n^{}$, $S_L^p$ and $S_L^h$ are functions of the scattering coefficients in Eqs.~\eqref{eq:resmatchingL}-\eqref{eq:scatbarrier}, for which we obtain a closed analytical expression, see Appendix~\ref{sec:recur}. 

\subsection{Calculation of the full current}\label{sec:numerics}

In this section, we briefly summarize the calculation of the current across the junction. The current [Eq.~\eqref{eq:currentdiff}] consists of contributions from states inserted from either reservoir, which are described by a set of reservoir parameters (i.e., the chemical potentials $\mu_L$ or $\mu_R$ and the gaps $\Delta_L$ or $\Delta_R$). Each contribution is defined in terms of an energy integral [Eq.~\eqref{eq:Idc}] that is evaluated numerically, where the integrand given in Eq.~\eqref{eq:Tpzeta} depends on the scattering coefficients of the Lippmann-Schwinger states (which are $\{a_{L/R,n},b_{L/R,n},c_{L/R,n},d_{L/R,n}\}$). For a given energy, these coefficients are obtained by first solving the recurrence relation [Eq.~\eqref{eq:recurrence}] to obtain the set $\{d_{L,n}\}$. For this infinite system of equations, efficient numerical solution algorithms exist in the form of the modified Lenz method~\cite{press02,hurd97} (see Appendix~\ref{sec:Lenz} for details). Remaining scattering coefficients are then obtained by substituting back into Eqs.~\eqref{eq:resmatchingL}--\eqref{eq:scatbarrier}. For quasiparticles injected from the right reservoir, we use the same approach to solve for the scattering coefficients (see Appendix~\ref{sec:recur}).

\subsection{Tunneling current}\label{sec:tunneling}

An important check of our calculations irrespective of the Andreev approximation is made by taking the tunneling limit: For tunnel junctions with small transparency \mbox{$t^2 \ll 1$} (i.e., \mbox{$|t_{p,n}| = |t_{h,n}| = t\to 0$}), the current arises due to direct transmission from the occupied states in one reservoir to empty states of the other reservoir. Standard calculations using a tunneling Hamiltonian predict a current~\cite{mahan00}
\begin{equation}
I_{\mathrm{dc}}(V) = 
\frac{2}{h}t^2\int_{-\infty}^{\infty} dE \, \rho_{L}(E) \rho_{R}(E+V) [f(E) - f(E+V)] , \label{eq:tunnelling}
\end{equation}
which is proportional to the product of the particle tunneling densities of state $\rho_{L/R}$ in both reservoirs. Since the tunneling current is due to direct transmission instead of multiple Andreev reflections, it can flow only if the voltage is greater than the energy difference between the occupied band of one reservoir and the empty band of the other reservoir, i.e., if $|V| \geq \bar{\Delta}_L + \bar{\Delta}_R$ where $\bar{\Delta}_{L/R}$ is the spectral gap of the fermion dispersion ($\bar{\Delta}= \Delta$ for $\mu > 0$ and $\bar{\Delta} = [\mu^2+\Delta^2]^{1/2}$ for $\mu \leq 0$; see Fig.~\ref{fig:1}). Note that since the Landauer-B\"uttiker formalism does not take into account the density of states in the final reservoir, this is a  strong check of our results.

%++++++++++++++++++++++++++++++++++++++++
\begin{figure}[t]
\includegraphics[width=\linewidth]{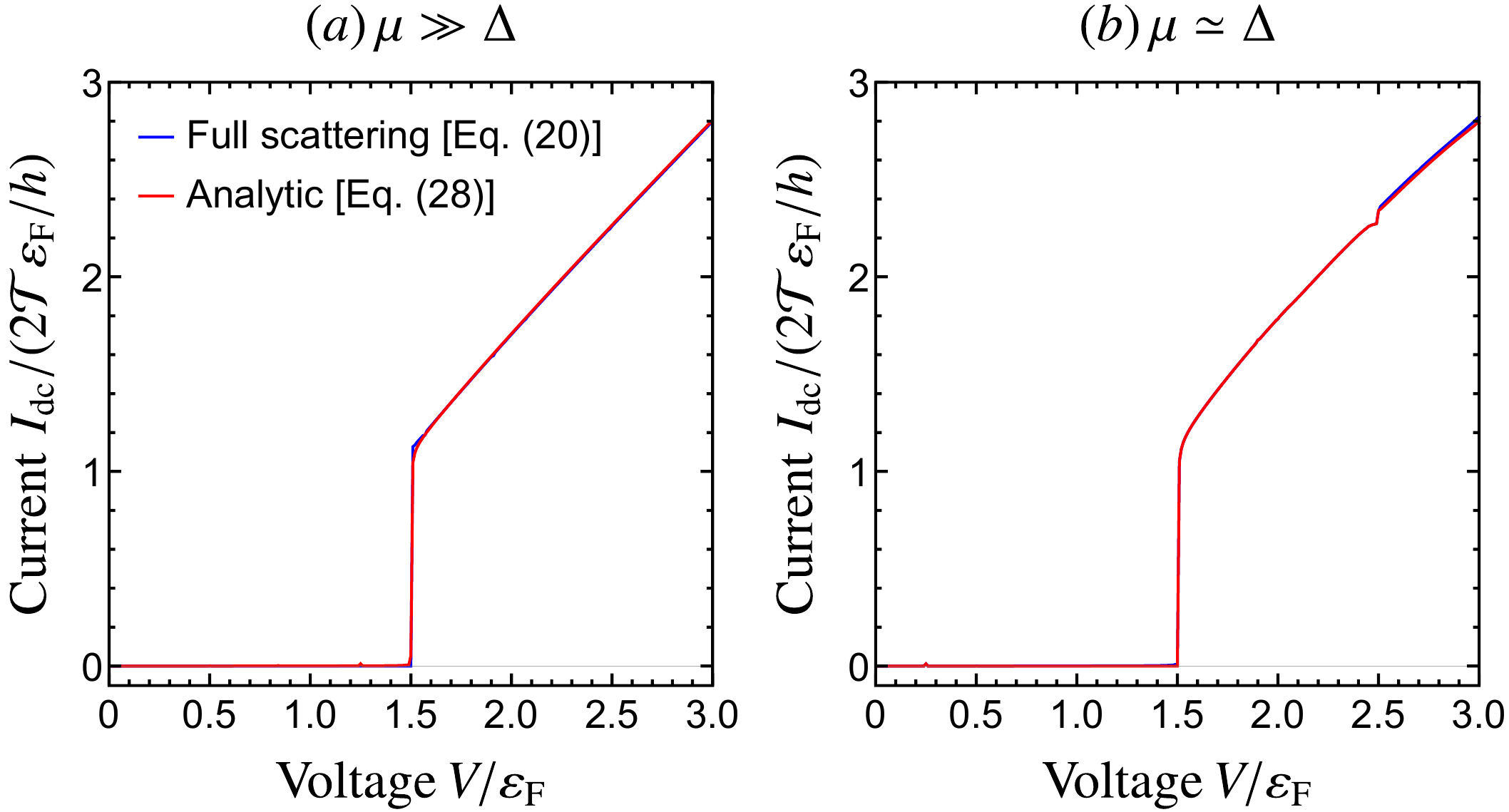}
\caption{
Current $I_{\mathrm{dc}}$ vs voltage $V$ calculated using the full scattering formula [blue lines; Eq.~\eqref{eq:Idc}] and the analytical tunneling expression [red lines; Eq.~\eqref{eq:tunnelling}], demonstrating excellent agreement between the two. Shown are two cases where (a) the Andreev approximation holds ($\mu_L/\EF = 100$) and (b) the Andreev approximation does not hold ($\mu_L/\EF = 3$). 
The tunneling current can flow only if $V \geq \Delta_L + \Delta_R = 1.5 \varepsilon_F$.
}\label{fig:check}
\end{figure}
%++++++++++++++++++++++++++++++++++++++++

Equation~\eqref{eq:tunnelling} is derived analytically within our framework by taking the tunneling limit of the full current in Eq.~\eqref{eq:Idc}. In this limit, the reflection and transmission coefficients across the delta-function barrier, which are given in Eqs.~\eqref{eq:tpn} and~\eqref{eq:rpn}, reduce to $r_{p,n} = r_{h,n}^* = -(1 - t^2/2) + \mathcal{O}(t^4)$ and $t_{p,n} = t_{h,n}^* = -it$ corresponding to $\eta = -\pi/2$. This implies that terms of order ${\cal O}(t^n)$ in the scattering coefficients represent transmission processes across the tunnel barrier of at least $n$ times. For $t^2 \ll 1$, we only need to consider terms up to ${\cal O}(t^2)$, which limits the number of contributing scattering channels to \mbox{$|n| \leq 2$}; see Fig.~\ref{fig:scatteringstates}. Substituting these results into the expression [Eq.~\eqref{eq:Tpzeta}] for the tunneling density, which is used to evaluate the current [Eqs.~\eqref{eq:currentdiff} and~\eqref{eq:Idc}], we obtain the tunneling current as given in Eq.~\eqref{eq:tunnelling}. The full derivation is detailed in Appendix~\ref{app:tunneling}. 

Figure~\ref{fig:check} shows the current calculated using the analytic tunneling formula [Eq.~\eqref{eq:tunnelling}] (red lines) and the full expression [Eq.~\eqref{eq:Idc}] (blue lines). We show plots for two separate cases: Figure~\ref{fig:check}(a) is for a case where the Andreev approximation holds ($\mu_L/\EF = 100$) and Fig.~\ref{fig:check}(b) represents a case where the Andreev approximation does not apply ($\mu_L/\EF = 3$). The parameters used for both plots are $\Delta_L/\EF = 1$, $\Delta_R/\EF = 0.5$, and $\mathcal{T} = 0.001$. Note that the kink in Fig.~\ref{fig:check}(b) corresponds to the transmission of particles from the van Hove singularities of the left normal region (\mbox{$E = -\mu_L = -3\EF$}) into a quasihole at the gap edge of the right reservoir (\mbox{$E = -\Delta_R = -0.5\EF$}); this tunneling process occurs at a voltage $V/\EF = (\mu_L -\Delta_R)/\EF = 2.5$. In both cases, there is excellent agreement, thus demonstrating that the full calculation reduces to the tunneling expression in Eq.~\eqref{eq:tunnelling} for small junction transparencies.

\section{Josephson junctions in the BCS-BEC crossover}\label{sec:results}

Having established a general Landauer-B\"uttiker framework for transport across a Josephson junction that does not rely on the Andreev approximation, we proceed in this section to apply our general formalism to superconducting junctions along the BCS-BEC crossover. We first outline and present the main transport results for different regimes along the crossover, and then proceed to discuss in detail the emergence of negative differential conductance at unitarity in Sec.~\ref{sec:negativeconductance} as well as the suppression of the subgap current as the current is tuned from the BCS to the BEC regime in Sec.~\ref{sec:splitting}.

A Josephson junction in the BCS-BEC crossover is described by three parameters: First, the density imbalance between reservoirs,
\begin{align}
\nu = \frac{n_L - n_R}{n_L + n_R} , \label{eq:imbalance}
\end{align}
which sets the voltage bias \mbox{$V = \mul - \mur$} that induces a particle current across the junction; second, the interaction strength 
\begin{align}
\frac{1}{k_F a} \label{eq:interactionstrength}
\end{align}
where $a$ is a three-dimensional scattering length and the Fermi momentum $\kF = (3\pi^2 \bar{n})^{1/3}$ is defined in terms of the average density of both reservoirs, \mbox{$\bar{n} =(n_L + n_R)/2$}, with a corresponding unit of energy $\EF = \hbar^2\kF^2/2m$; and third, the tunnel-barrier transparency $\mathcal{T}$ that sets the strength of the delta-function scatterer in Eqs.~\eqref{eq:tpn} and~\eqref{eq:rpn}. The BCS-BEC crossover is tuned by the interaction strength $1/k_F a$, which is either done by varying the carrier density (with high densities corresponding to small interactions strengths), or by changing the scattering length as done in cold atom experiments. 

%++++++++++++++++++++++++++++++++++++++++
\begin{figure}[t]
\includegraphics[width=\linewidth]{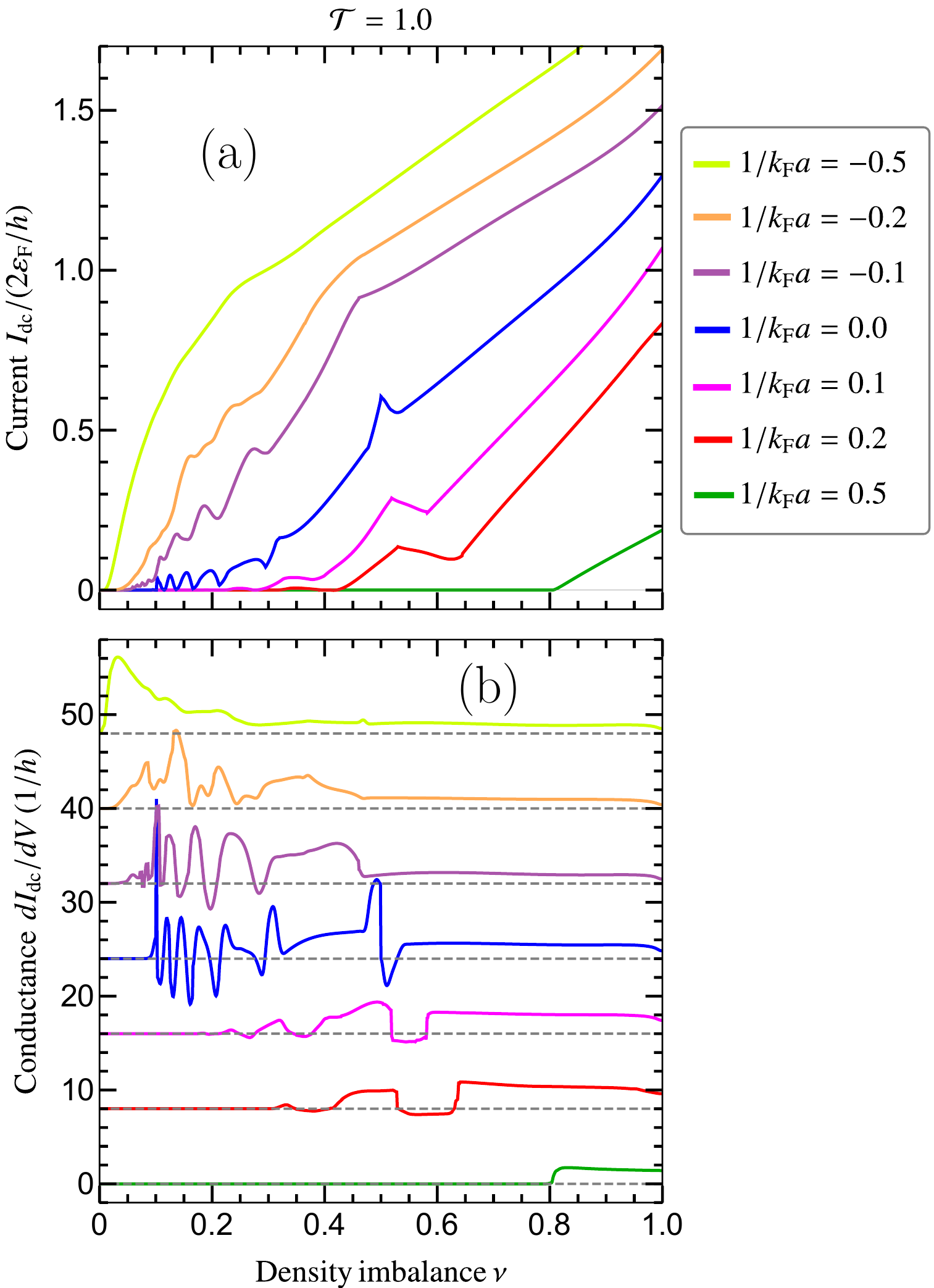}
\caption{(a) DC current $\Idc$ and (b) differential conductance $dI_{\mathrm{dc}}/dV$ of a transparent junction ($\mathcal{T} = 1$) as a function of density imbalance $\nu$ along the BCS-BEC crossover (each differential conductance curve is shifted by $8/h$). Close to unitarity \mbox{$1/\kF a = 0$}, the current becomes strongly nonlinear with sharp peaks and dips resulting in negative differential conductance at specific voltages. The chemical potentials and pairing gaps as the inputs to calculate the current and conductance are shown in Fig.~\ref{fig:spectral}.
}
\label{fig:2}
\end{figure}
%++++++++++++++++++++++++++++++++++++++++

%++++++++++++++++++++++++++++++++++++++++
\begin{figure*}[t]
\includegraphics[width=\linewidth]{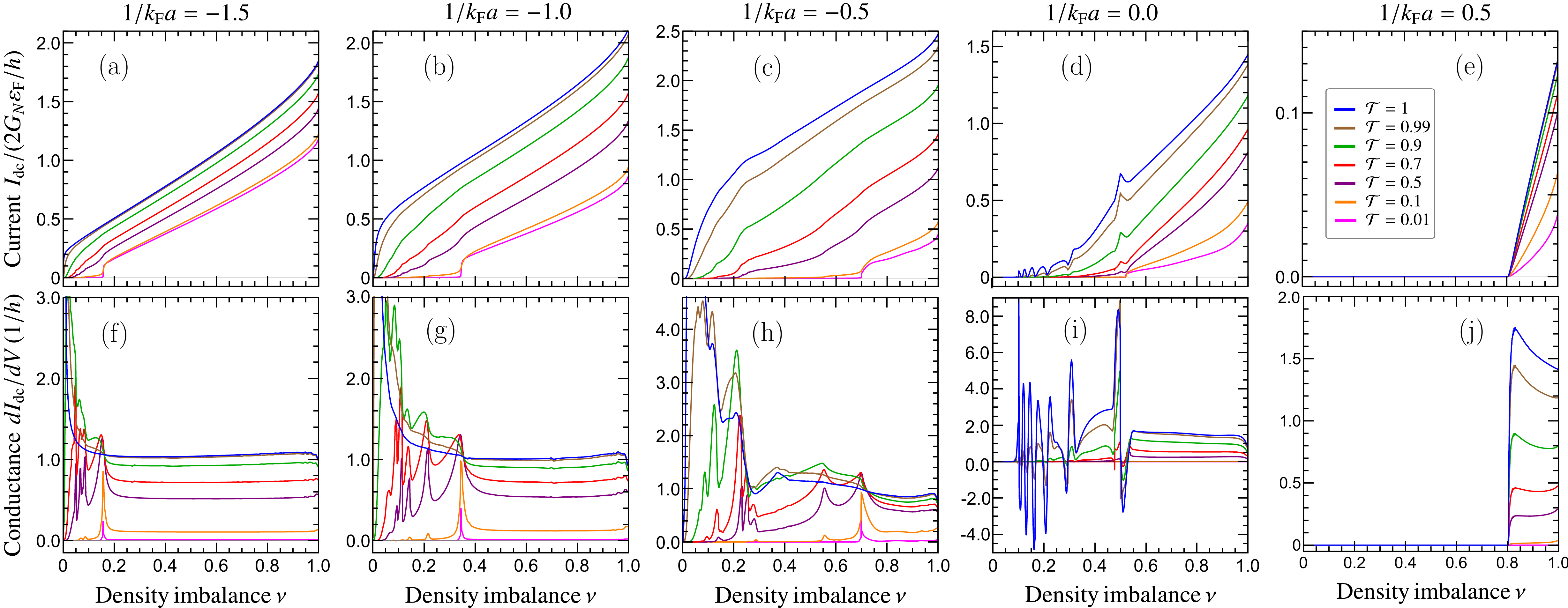}
\caption{DC current $\Idc$ (upper panels) and differential conductance $d\Idc/dV$ (lower panels) as a function of the density imbalance~$\nu$ between reservoirs~$\nu$ for different transparencies $\mathcal{T}$ as the system is tuned from the BCS to the BEC regime (left to right). Upper panels show the current $\Idc$ normalized by the dimensionless conductance at $\nu = 1$, $G_N \equiv h \times dI_{\mathrm{dc}}/dV(\nu=1)$. The chemical potentials and pairing gaps used as the inputs to calculate the current and conductance are shown in Fig.~\ref{fig:spectral}.}
\label{fig:3}
\end{figure*}
%+++++++++++++++++++++++++++++++++++++++

For small attractive interactions (\mbox{$1/k_F a \to - \infty$}), the reservoirs are in the BCS limit, where the chemical potential vastly exceeds the pairing gap in either reservoir ($\mu \gg \Delta$). This is the regime of conventional superconductivity where the Andreev approximation is expected to hold. Here, we expect and confirm conventional transport signatures, where a subharmonic gap structure appears due to multiple Andreev reflections with a weak onset resonance whenever the voltage bias matches a harmonic fraction of the pairing gaps in the reservoirs. As the interaction strength increases but still in the attractive regime (\mbox{$1/k_F a < 0$}), the magnitude of the pairing gap increases and becomes comparable to the chemical potential. As we will demonstrate, already in this regime of mild deviation from the strict BCS limit, the Andreev approximation is no longer valid and normal reflections start to play an important role. In addition, the transport is now also affected by the van Hove singularities in the band structure, which become lower in energy as the junction is tuned away from BCS limit [see Fig.~\ref{fig:1}(c)]. These van Hove singularities manifest as new features in the subharmonic gap structure. 

At unitarity (\mbox{$1/k_F a = 0$}), the interaction changes sign where the infinitely attractive and repulsive limit are continuously connected. This point marks the crossover from the BCS to the BEC regime. Here, the Andreev approximation is completely invalid, and there is a strong enhancement of normal reflection processes. Indeed, as we show, the subharmonic gap structure takes a complex form marked by even {\it negative} differential conductance whenever the voltage bias matches a harmonic fraction of the van Hove energies. 

Finally, as the interaction is tuned to the repulsive regime (\mbox{$1/k_F a > 0$}), the pairing gap increases further while the chemical potential decreases rapidly and becomes negative for sufficiently strong repulsive interaction. There is a particular interaction strength on the BEC side (known as the splitting point) at which the position of the minimum of the superconducting band structure changes from finite to zero momentum [illustrated in the lower panel of Fig.~\ref{fig:1}(c)]. Beyond this point, multiple Andreev reflections are no longer possible, leading to a complete suppression of the subgap current. The splitting point plays an important role in the phase diagram of spin-imbalanced Fermi gases~\cite{son06,haussmann09,frank18}, where it separates unpolarized superfluid phases, gapless superfluids, and Fulde-Ferrell-Larkin-Ovchinnikov phases. Transport measurements in a Josephson junction transport offer a way to detect this point experimentally. 

We note that all transport signatures that emerge beyond the Andreev approximation, such as the negative differential conductance at unitarity and the suppression of the Andreev current as the interaction is tuned to the BEC regime, are linked to the superconducting band structure. As such, it is crucial to take into account the energy-dependence of various Andreev and normal reflections processes, and it is not possible to obtain these features using standard Andreev approximation calculations~\cite{averin95,hurd96,hurd97,octavio83,flensberg88,furusaki92} combined with phenomenological approaches, such as including additional tunnel barriers~\cite{octavio83} or a Fermi velocity mismatch~\cite{Mortensen1999Angle}.

In our calculations, we relate $a$ and $\nu$ to the reservoir chemical potentials and pairing gaps using the bulk mean-field equations, which are summarized in Appendix~\ref{sec:bulkmeanfield}. This choice is not specific to our method and these parameters can also be taken from full many-body calculations~\cite{frank18}. Note that while we present results in terms of the density imbalance instead of the bias voltage, both quantities are proportional for small $\nu$ with
\begin{align}\label{eq:VEF}
\frac{V}{\EF}= \frac{(\mu_L -\mu_R)}{\EF} = \frac{2}{\kappa \EF \bar{n}} \nu,
\end{align}
where 
$\kappa \equiv (1/\bar{n}^2) \times (\partial \bar{n}/\partial\mu)$ is the electronic compressibility. For mean field parameter values, the proportionality holds for nearly all density imbalances as illustrated in Appendix~\ref{sec:bulkmeanfield}.

We proceed in the remainder of this section to present comprehensive results of the dc current across the Josephson junction for all interaction strengths and transparencies, and then discuss detailed features in subsequent sections. Figure~\ref{fig:2} shows the dc current (top panel) and differential conductance (bottom panel) for different scattering lengths along the BCS-BEC crossover, calculated  for perfect transparency \mbox{$\mathcal{T} = 1$},
as a function of density imbalance $\nu$. In the strict BCS limit (\mbox{$1/k_Fa\ll0$}, light green line in Fig.~\ref{fig:2}), multiple Andreev reflections give rise to a large subgap current with nonanalyticities at voltages given by harmonic fractions of the superconducting gap, consistent with results in Refs.~\cite{klapwijk82,octavio83,arnold87,gunsenheimer94,averin95,hurd96,cuevas96,hurd97}. As the interaction is tuned toward the BEC limit, the current decreases, indicating that the reservoirs become more insulating where the strength of normal and Andreev reflections increases and decreases, respectively. Near unitarity (\mbox{$1/k_Fa \to 0$}), the subgap current develops sharp dips giving rise to negative differential conductance, which is an experimental signature of the unitary regime. This is our first central result, and it is discussed in detail in Sec.~\ref{sec:negativeconductance}. 

Further on the BEC side of the crossover (dark green lines in Fig.~\ref{fig:2}), the multiple-Andreev-reflection induced subgap current is completely suppressed below a threshold imbalance. This complete suppression happens as the interaction is tuned to the BEC limit (at and beyond the splitting point) and will be discussed in detail in Sec.~\ref{sec:splitting}. The residual current above threshold imbalance is linked to a single Andreev reflection process. As discussed above, these features are not artifacts of the sharp gap and chemical potential profiles considered here, but are linked to the van Hove points in the superconducting band structure at which normal reflections are significantly enhanced.

%++++++++++++++++++++++++++++++++++++++++
\begin{figure*}[t]
\includegraphics[width=\linewidth]{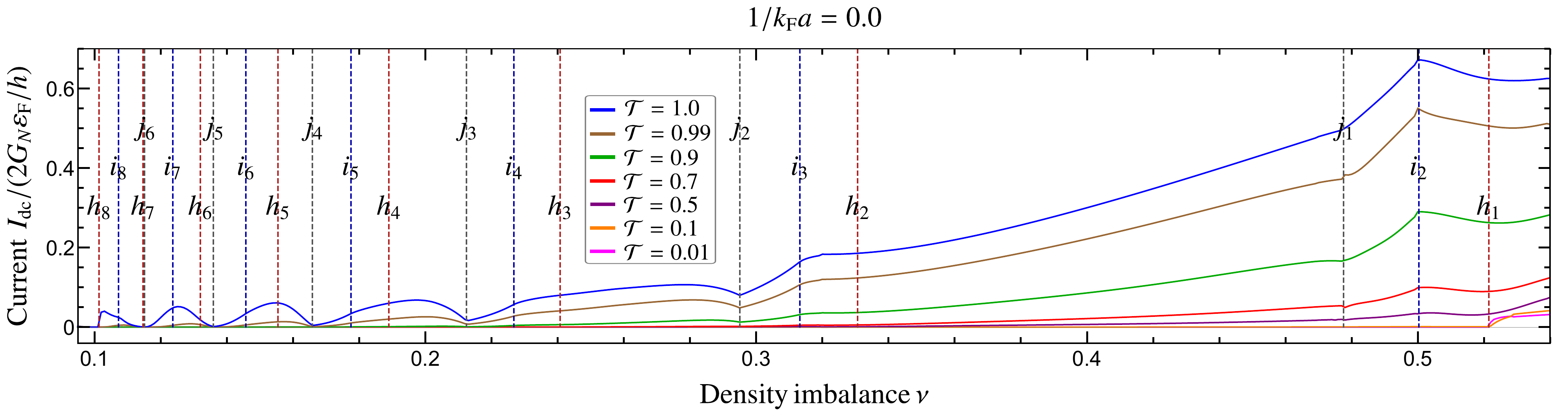}
\caption{
Subharmonic gap structure at unitarity ($1/\kF a = 0$). The current is the same as in Fig.~\ref{fig:3}(d). Vertical dashed lines denote the density imbalance at which the subharmonic gap structure occurs, with corresponding voltages given in Table~\ref{table:SGS} of Appendix~\ref{sec:SGS}. The chemical potentials and pairing gaps used as the inputs to calculate the current are shown in Figs.~\ref{fig:spectral}(II.d) and~\ref{fig:spectral}(II.i).
}\label{fig:SGS}
\end{figure*}
%++++++++++++++++++++++++++++++++++++++++

%++++++++++++++++++++++++++++++++++++++++
\begin{figure*}[t]
\includegraphics[width=\linewidth]{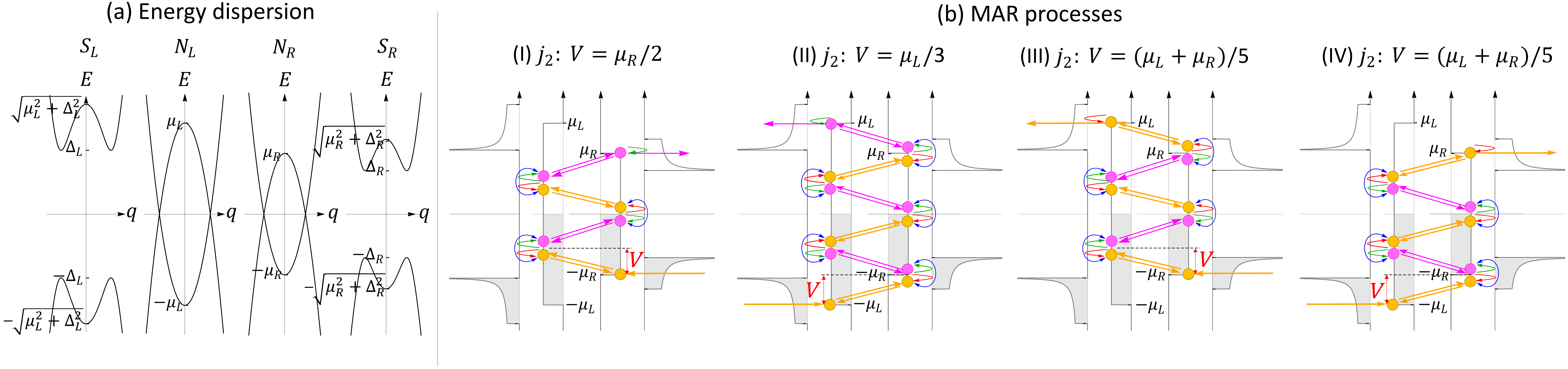}
\caption{
(a) Sketch of the energy dispersion of a superconducting junction at unitarity, where the chemical potentials of both reservoirs are positive and comparable to the superconducting gap. (b) Four different multiple Andreev reflection (MAR) processes that contribute to the point labeled $j_2$ in Fig.~\ref{fig:SGS}, which appears as a dip in the current at a voltage $V = \mu_R/2=\mu_L/3 = (\mu_L+\mu_R)/5$. The reduction in the current is linked to the enhanced normal reflection at the bottom of the particle band and the top of the hole band as shown in Fig.~\ref{fig:S4}. Shown are the Andreev reflection (blue arrow), and the normal reflection of particles (red arrow) and holes (green arrow). 
}\label{fig:MARIm}
\end{figure*}
%++++++++++++++++++++++++++++++++++++++++

Besides tuning the interaction from the BCS to BEC regime, the subgap current is also suppressed by decreasing the barrier transparency. To compare these effects, we show in Fig.~\ref{fig:3} the current $\Idc$ (upper panels) and differential conductance (lower panels) as a function of $\nu$ along the BCS-BEC crossover (left to right panels) for different barrier transparencies, interpolating between the tunneling limit (\mbox{${\cal T} \to 0$}), where it matches Eq.~\eqref{eq:tunnelling}, and the perfectly transparent barrier (\mbox{${\cal T} \to 1$}) shown in Fig.~\ref{fig:2}. The current generally decreases with decreasing barrier transparency $\mathcal{T}$. 
 The current-voltage dependence in the BCS regime becomes more nonlinear at intermediate transparencies with kinks in the current and resonances in the differential conductance~\cite{arnold87,averin95,cuevas96}. However, there is no negative differential conductance, which is thus visibly different from the unitarity limit shown in Figs.~\ref{fig:2},~\ref{fig:3}(d), and~\ref{fig:3}(i).  As shown in Figs.~\ref{fig:3}(d) and~\ref{fig:3}(i), the subharmonic gap structure at unitarity is still visible for intermediate values of junction transparency ($\mathcal{T} \gtrsim 0.7$), implying that this feature is quite robust against disorder.  On the BEC side [Figs.~\ref{fig:3}(e) and~\ref{fig:3}(j)], the subgap current due to multiple Andreev reflections vanishes beyond the splitting point, even for the fully transparent junction, with a concomitant reduction in the single-Andreev reflection induced current. 

%++++++++++++++++++++++++++++++++++++++++
\begin{figure*}[t]
\includegraphics[width=1.\linewidth]{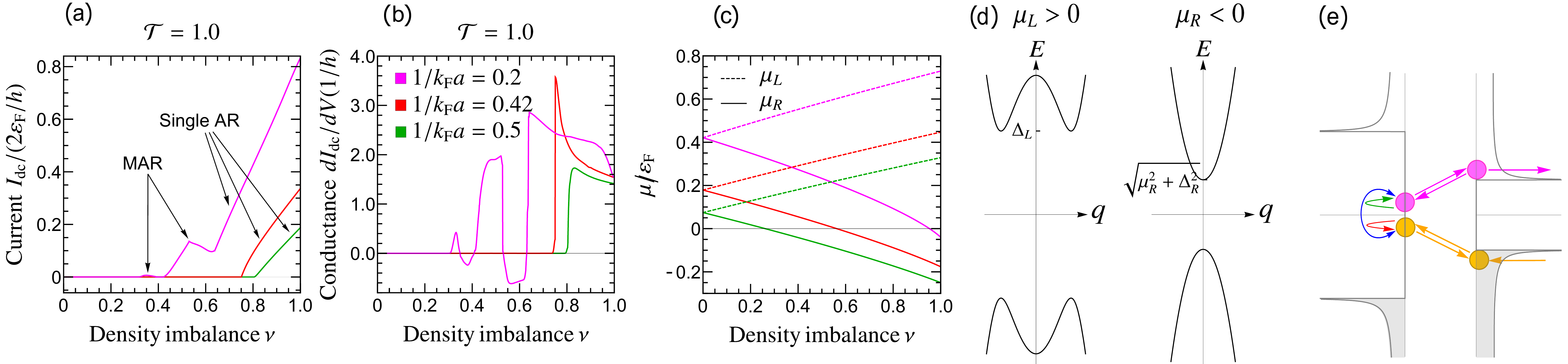}
\caption{
(a) DC current $\Idc$, (b) differential conductance $dI_{\mathrm{dc}}/dV$,  and (c) chemical potential of the left ($\mu_L$, dashed) and right reservoir ($\mu_R$, solid) for different interaction strengths $1/\kF a$ near the splitting point (shown in different colors) as a function of the density imbalance $\nu$. The current and differential conductance are calculated for a transparent junction ($\mathcal{T} = 1$) and the chemical potentials are plotted in unit of the Fermi energy $\EF$. For the interaction strength at and above the splitting point ($1/k_F a \gtrsim 0.42$), current flows only via a single Andreev reflection (AR) at the left normal-superfluid interface and the subgap current due to multiple Andreev reflections (MAR) is suppressed due to the absence of Andreev reflection at the right normal-superconductor interface. (d) At the splitting point, the chemical potential of the right superfluid ($\mu_R$) becomes negative, which (within mean-field theory) changes the curvature of its quasiparticle dispersion (i.e., its minimum is shifted from finite to zero momentum). (e) Schematics of the single Andreev reflection process at the left normal-superfluid interface that gives rise to the current at and above the splitting point. Shown are the Andreev reflection (blue arrow), and the normal reflection of particles (red arrow) and holes (green arrow). 
}\label{fig:splitting}
\end{figure*}
%++++++++++++++++++++++++++++++++++++++++

\subsection{Negative differential conductance as a signature of BCS-BEC crossover}\label{sec:negativeconductance}

In this section, we discuss the subharmonic gap structure---i.e., the nonanalyticities in the dc current as a function of density imbalance---that is visible in the dc current shown in Figs.~\ref{fig:2} and~\ref{fig:3}, focusing in particular on features beyond the Andreev approximation. 
Within the Andreev approximation, it is understood that these nonanalyticities occur whenever the voltage matches a harmonic fraction of the reservoir gaps, which marks the onset of an additional Andreev reflection channel. Our results beyond the Andreev approximation show additional subgap harmonic structures in the BCS-BEC crossover. These new subgap features arise as a result of multiple Andreev reflections in conjunction with the enhanced normal reflection at the van Hove points in the superconducting band structure. They appear as a reduction (or dip) in the current (and hence negative differential conductance) at specific voltages that corresponds to a harmonic series of the reservoir chemical potential. 
In Fig.~\ref{fig:SGS}, which is explained below, we show quantitatively that the voltages at which these subgap features appear are set by harmonic fractions of the van Hove singularities of the left ($E_{\mathrm{vHs},L}$) and right ($E_{\mathrm{vHs},R}$) regions of the superfluid-normal-superfluid junction. These van Hove singularities, which mark discontinuities in the density of states of the superfluid or normal regions in the junction, are located at energies $E_{\mathrm{vHs}} = \mu, \Delta$, or $\sqrt{\mu^2+\Delta^2}$. 

To illustrate this, we begin by considering a superconducting junction in which both reservoirs have positive chemical potentials ($\mu_{L}>0$ and $\mu_{R}>0$). The corresponding  band structure of the superconducting and normal regions in the junction is sketched in Fig.~\ref{fig:MARIm}(a). The subharmonic gap structure occurs at specific voltages that can be expressed as harmonic series of the energies of the van Hove singularities $E_{\mathrm{vHs},L}$ and $E_{\mathrm{vHs},R}$: 
\begin{subequations}\label{eq:SGS}
\begin{align}
 |V| &= E_{\mathrm{vHs},L}/n,\label{eq:SGSa}\\
|V| &= E_{\mathrm{vHs},R}/n,\label{eq:SGSb}\\
|V| &= (E_{\mathrm{vHs},L} + E_{\mathrm{vHs},R})/(2n-1)\label{eq:SGSc}
\end{align}
\end{subequations}
 where $n \in \mathbb{Z}^+$ is a positive integer. First, the subharmonic gap voltages in Eq.~\eqref{eq:SGSa} correspond to the case where a quasiparticle enters the junction from the occupied band of the left superfluid and undergoes $(2n-1)$ Andreev reflections until it is transmitted into the empty band of the left superfluid. Second, the voltages in Eq.~\eqref{eq:SGSb} correspond to a similar case where a quasiparticle enters the junction from the right superfluid, undergoes $(2n - 1)$ Andreev reflections until it is transmitted into the empty band of the right superfluid. Third, the subharmonic gap voltages in Eq.~\eqref{eq:SGSc} corresponds to the case where a quasiparticle entering the junction from either the left/right superfluid, undergoes $(2n-1)$ Andreev reflections until it is transmitted into the empty band of the opposite reservoir. In the following, we show that Eq.~\eqref{eq:SGS} indeed quantitatively describes the position of the subgap features, and then proceed to illustrate the various processes in more detail.

Figure~\ref{fig:SGS} shows a magnified plot of the dc current of Fig.~\ref{fig:3}(d) at unitarity ($1/\kF a = 0$), where $\mu_L \simeq \Delta_{L} > 0$ and $\mu_R \simeq \Delta_{R} > 0$. We indicate using dashed lines the positions of the subharmonic gap structure that are visible in the current, where the labels of these subgap features with their corresponding voltages are given by
\begin{subequations}\label{eq:Vm}
\begin{align}
h_n: V_n &= \Delta_R/n,\\
i_n: V_n &= (\mu_L+\Delta_R)/(2n-1),\\
j_n: V_n &= \mu_R/n,
\end{align}
\end{subequations}
where $n \in \mathbb{Z}^+$ is a positive integer. Note that in this paper, we show the subharmonic gap structure only for positive voltages corresponding to $n \in \mathbb{Z}^+$ even though Eqs.\eqref{eq:SGS} and~\eqref{eq:Vm} also hold for negative voltages. The positions are determined using Eq.~\eqref{eq:SGS}, which takes only the mean field chemical potentials and pairing gaps as inputs. Corresponding numerical values for the positions are listed in Appendix~\ref{sec:SGS}. The very precise agreement between the position of the subharmonic gap structure obtained from our transport calculation and Eq.~\eqref{eq:SGS}, which is done independently of our transport calculation, is an extremely strong check of our results and confirms the underlying physics that links these features to the superconducting band structure. Of particular importance are the subharmonic-gap-structure positions labeled by $j_m$, for which dips appear in the current that give rise to negative differential conductance. This subharmonic gap structure appears at voltages $V_n = \mu_{R}/n = \mu_{L}/(n+1) = (\mu_L+\mu_{R})/(2n+1)$, where the second and third equalities follow from the definition $V = \mu_L - \mu_R$. At these voltages, multiple Andreev reflections send particles from the band bottom of the particle dispersion $(E = -\mu)$ to the band top of the hole dispersion ($E=\mu$) of the same or opposite region of the superconducting junction [see Fig.~\ref{fig:MARIm}(b)]. 

Let us discuss the physical mechanism that give rise to negative differential conductance in more detail. In Fig.~\ref{fig:MARIm}(b), we show all multiple Andreev reflection processes that contribute to one particular current dip point $j_2$ that is located at $\nu=0.295$ [see~Fig.~\ref{fig:SGS}]. Since the band bottom (top) of the particle (hole) dispersion marks the energy beyond which the particle (hole) density of states vanishes, we have perfect normal reflections for particles (holes) and completely suppressed Andreev reflections close to and beyond these van Hove points. As a result, the multiple Andreev reflections that link these two van Hove singularities are suppressed and give rise to dips in the current and hence negative differential conductance. 
We note that if instead of taking $\mu_{N,L} = \mu_{S,L} = \mu_L$ and $\mu_{N,R} = \mu_{S,R} = \mu_R$ for the chemical potentials in the normal region (as in this paper), we chose $\mu_{N,L} > [\mu_{S,L}^2+\Delta_{L}^2]^{1/2}$ and $\mu_{N,R} > [\mu_{S,R}^2+\Delta_{R}^2]^{1/2}$, i.e., where the band bottom (top) of the particle (hole) dispersion in the normal is at an energy below (above) those in the superfluid, the dips in the current would occur at $V_n = [\mu_{S,R}^2+\Delta_{R}^2]^{1/2}/n$, $[\mu_{S,L}^2+\Delta_{L}^2]^{1/2}/(n+1)$, and $([\mu_{S,L}^2+\Delta_{L}^2]^{1/2}+[\mu_{S,R}^2+\Delta_{R}^2)]^{1/2}/(2n+1)$, which would lead to a small quantitative shift in the subharmonic gap structure but no qualitative change. 

Negative differential conductance appears routinely in mesoscopic devices whenever the bias voltage is tuned to allow resonant tunneling to bound states in the junction, see, for example, Refs.~\cite{Yeyati1997Resonant,zhitlukhina2016anomalous,Huang2021Spin,san2013multiple}. Note that this resonant bound-state tunneling is different from our mechanism for negative differential conductance: First, unlike the physical-bound-state case, our negative differential conductance will disappear as the interaction is detuned from unitarity. Second, for our mechanism, the differential conductance measured in the tunneling limit will not have any bound-state tunneling peaks. 

\subsection{Suppression of multiple Andreev reflections at the splitting point}\label{sec:splitting}

As a second main result, we propose an experimental protocol to determine the splitting point, namely by measuring the current or differential conductance as the interaction is tuned into the BEC ($1/k_Fa\gg0$) regime, where the splitting point defines the critical interaction strength at which the subgap current or differential conductance due multiple Andreev reflection vanishes [see Fig.~\ref{fig:splitting}(a) and~\ref{fig:splitting}(b)]. This is because at this critical interaction strength, the chemical potential of the right reservoir turns negative (i.e., the quasiparticle-dispersion curvature changes). Beyond this point multiple Andreev reflections are suppressed and current flows only due to a single Andreev reflection at the left normal-superfluid interface. This is explained in more detail in the following discussion. 

An Andreev reflection process requires overlapping particle and hole bands in the superconducting band structure, which only exist for a positive chemical potential. This is illustrated for the band structure in Figs.~\ref{fig:1}(c) and~\ref{fig:MARIm}(a) and is shown quantitatively in Fig.~\ref{fig:S4}. Correspondingly, as soon as the chemical potential of a reservoir turns negative, an Andreev reflection at the corresponding normal-superconductor interface can no longer take place, hence the vanishing of the subgap current as the interaction strength is tuned to the BEC side of the crossover [see Figs.~\ref{fig:splitting}(a) and~\ref{fig:splitting}(b)]. Above a critical interaction strength, the chemical potential of the right reservoir turns negative [Fig.~\ref{fig:splitting}(c)], and its quasiparticle dispersion changes curvature [Fig.~\ref{fig:splitting}(d)], which (for mean-field parameters) defines the splitting point. Note that the residual current at and beyond the splitting point is not due to direct tunneling, but due to a single Andreev reflection, which occurs at the left normal-superfluid interface where the chemical potential is still positive [see Fig.~\ref{fig:splitting}(d)]. This results in an Andreev current for voltages $|V| \geq \sqrt{\mu_R^2+\Delta_R^2}$ [Fig.~\ref{fig:splitting}(e)].  Even though there is a subgap current due to a single Andreev reflection, the subharmonic gap structure due to multiple Andreev reflections is completely suppressed. Only far on the BEC side of the crossover, where the chemical potentials of both reservoirs become negative, is this single Andreev reflection process completely suppressed, which then results in no current at all. 

This characteristic suppression of the current on the BEC side allows us to identify the splitting point as the critical interaction strength at which the subgap current due to multiple Andreev reflections vanishes as the system is tuned towards the BEC limit.  Our calculation using mean-field parameters shows that this point occurs at an interaction strength of $1/\kF a \approx 0.42$. Calculations beyond mean-field theory predict a value \mbox{$1/k_Fa \approx 0.8$} further on the BEC side~\cite{haussmann09,frank18}, implying that the current on the BEC side is more pronounced for non-mean-field parameters. 

\section{Conclusions}\label{sec:conclusions}

In summary, we have presented a general transport framework for voltage-biased superconducting junctions without relying on the Andreev approximation. Our formalism can be used to describe transport in most systems of current interest where the Andreev approximation, which assumes a reservoir chemical potential much larger than the superconducting gap, is not valid. We apply this framework to provide a comprehensive discussion of transport across superconducting or superfluid junctions with $s$-wave interactions along the BCS-BEC crossover. Crucially, our general formalism reveals several transport features that cannot be captured using the Andreev approximation, in particular, negative differential conductance in the unitary regime and suppression of the subgap current on the BEC side of the crossover.

On a technical side, relaxing the Andreev approximation complicates further the already daunting complexity of existing calculations that use this approximation~\cite{averin95,hurd96,hurd97,arnold87,gunsenheimer94,cuevas96,bolech05}. We summarize three main differences between our formalism and a treatment that assumes the Andreev approximation:
\begin{enumerate}
\item In contrast to calculations using the Andreev approximation, where all quasiparticle momenta are energy-independent, i.e., assumed to be fixed at a (large) Fermi momentum \mbox{$k = k_F = \sqrt{2m\varepsilon_F}/\hbar$}, we take into account the full energy-dependence of the quasiparticle momenta.  
\item While the standard formalism using the Andreev approximation assumes that only perfect Andreev reflection (and no normal reflection) occurs at the normal-superfluid interfaces, we take into account normal reflections at the normal-superfluid interfaces. These normal reflections, particularly the perfect normal reflections for particles (holes) with energies below (above) the energy of the van Hove singularities of the particle (hole) band, give rise to new transport features that cannot be captured by assuming the Andreev approximation. 
\item In contrast to the Andreev approximation case, where quasiparticle entering from a reservoir can only be transmitted into the normal region as a particle and not as a hole, a quasiparticle entering from a reservoir can also be transmitted into the normal region as a hole, opening up an entirely new transmission channel. 
\end{enumerate}

Since transport features based on multiple Andreev reflections are routinely used to probe the spectral properties of superconducting junctions~\cite{Scheer1997Conduction,Buitelaar2003Multiple,Baer2014Experimental}, one can expect the negative differential conductance proposed here to be observed in experiments, especially in ultracold atomic junctions~\cite{valtolina15,husmann15,burchianti18,Xhani2020Critical,kwon2020strongly,luick2020ideal,pace21}, which are usually defect free. Indeed, the current across a superfluid point contact at unitarity has been measured in several quantum gas experiments~\cite{husmann15,krinner16}. There, the current at small voltage is larger than that in the BCS limit, which can be attributed to fluctuations and geometric effects in the reservoir~\cite{kanasznagy16,liu17,uchino17,krinner17,yao18} that are not modeled by our theory. Although the measured current at unitarity (Fig.~2(b) of Ref.~\cite{husmann15}) has an oscillatory form similar to our results, this appears to be a remnant of data processing~\cite{dsuppl}. The resolution of the nonlinear subgap current in quantum gas experiments, predicted by our theory as a signature of the BCS-BEC crossover, is thus an interesting prospect for future experiments. Our work also provides an experimental protocol to determine the splitting point, a central point in the cold-atom phase diagram. Furthermore, our formalism can be extended to calculate higher-harmonic (ac Josephson) currents, which are used to quantify the superfluid condensate fraction~\cite{meier01,Zaccanti2019Critical,kwon2020strongly}.

\begin{acknowledgments}
We thank Jean-Philippe Brantut, Sriram Ganeshan, Alejandro Lobos, Matteo Zaccanti, and Wilhelm Zwerger for comments and discussions. 
This work is supported by the Army Research Office Grant no. W911NF-19-1-0328 [F.S.] and Vetenskapsr{\r a}det (Grant No. 2020-04239) [J.H.].
\end{acknowledgments}

\appendix

\section{Bogoliubov states}\label{sec:bogoliubov}

In this appendix, we present the solution of the Bogoliubov equation for a constant gap profile, which describes excitations propagating in the $x$ direction deep in a single reservoir, i.e., where $\Delta(x)$ is independent of $x$ and equal to the bulk values $\Delta_L$ and $\Delta_R e^{i\phi}$, respectively. The explicit result for the wave function of an excitation with momentum $q$ along the $x$ direction is (recall that the prefactor is chosen such that for real $q$, the state has unit probability current)
\begin{align}
\Psi(q) &= \sqrt{\frac{m}{\hbar q}} \begin{pmatrix}
u(q) \\
v(q)
\end{pmatrix} e^{-i E_q \tau/\hbar} e^{i q x}, \label{eq:eigenstate}
\end{align}
with two energy branches (for an $s$-wave gap)
\begin{align}
E_q &= \pm \sqrt{(\varepsilon_q-\mu)^2 + |\Delta|^2}, \label{eq:dispersion}
\end{align}
where $\varepsilon_{q} = \frac{\hbar^2}{2m} q^2$, and the Bogoliubov coefficients are
\begin{subequations}\label{eq:uv}
\begin{align}
u^2(q) &= \frac{1}{2} \Bigl(1 + \frac{\varepsilon_{q} - \mu}{E_q}\Bigr), \\ 
v^2(q) &= \frac{1}{2} \Bigl(1 - \frac{\varepsilon_{q} - \mu}{E_q}\Bigr) .
\end{align}
\end{subequations}

%++++++++++++++++++++++++++++++++++++++++
\begin{figure}[t]
\includegraphics[width=\linewidth]{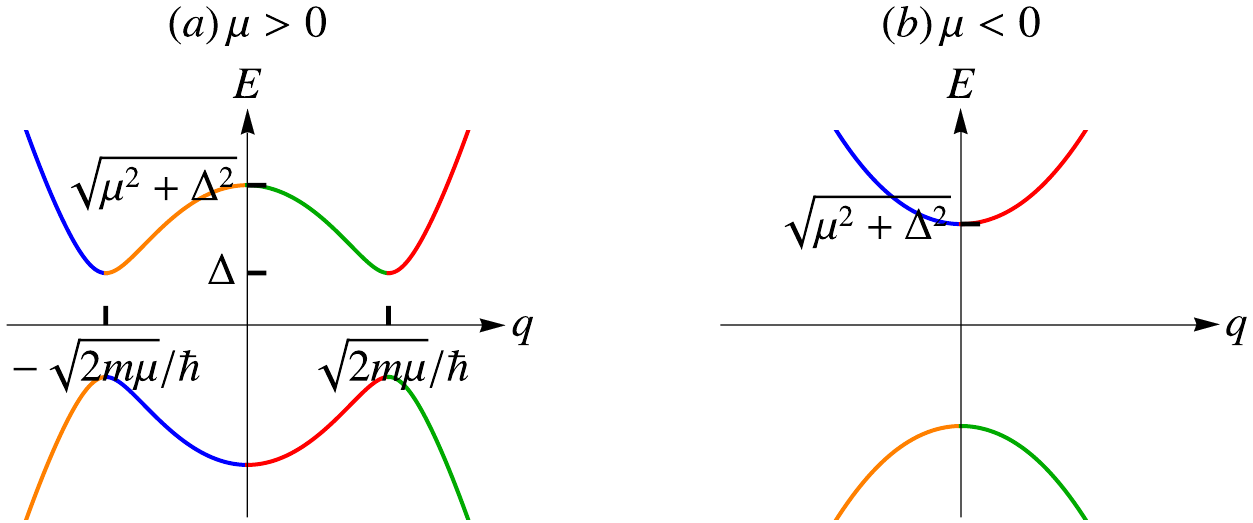}
\caption{Energy spectrum Eq.~\eqref{eq:dispersion} for (a) $\mu>0$ (corresponding to the BCS-side of the crossover) and (b) $\mu<0$ (corresponding to the BEC-side of the crossover).}
\label{fig:spectrum}
\end{figure}
%++++++++++++++++++++++++++++++++++++++++

The Bogoliubov spectrum~\eqref{eq:dispersion} is illustrated in Fig.~\ref{fig:spectrum} for (a) positive and (b) negative chemical potential.  We call states 
\begin{itemize}
\item {\it particle-like} if their group velocity $\tilde{v}(q) = (\partial E_q/\partial q)/\hbar$ has the same sign as $q$ (red and blue lines in Fig.~\ref{fig:spectrum}, orange lines in Fig.~\ref{fig:1} of the main text), and
\item {\it hole-like} if the group velocity and momentum have opposite sign (orange and green lines in Fig.~\ref{fig:spectrum}, magenta lines in Fig.~\ref{fig:1} of the main text). 
\end{itemize}
Furthermore, we denote a state as
\begin{itemize}
\item {\it right-moving} if it has positive group velocity (orange and red lines in Fig.~\ref{fig:spectrum}), and
\item {\it left-moving} if it has negative group velocity (blue and green lines in Fig.~\ref{fig:spectrum}).
\end{itemize}
For a given energy, the momentum of the right-moving particle-like excitation (red line in Fig.~\ref{fig:spectrum}) is
\begin{widetext}
\begin{align}\label{eq:qpapp}
q_{p}(E) &= \frac{\sqrt{2m}}{\hbar} \begin{cases}
i [\sqrt{E^2 - |\Delta|^2} - \mu]^{1/2}, & E < - \sqrt{\mu^2+|\Delta|^2}, \\
[\mu + i 0 - \sqrt{E^2 - |\Delta|^2}]^{1/2}, & - \sqrt{\mu^2+|\Delta|^2} < E < - |\Delta|, \\
[\mu + i \sqrt{|\Delta|^2 - E^2}]^{1/2}, & -|\Delta| < E < |\Delta|, \\
[\mu + i 0 + \sqrt{E^2 - |\Delta|^2}]^{1/2}, & |\Delta| < E < \sqrt{\mu^2+|\Delta|^2}, \\
[\mu + \sqrt{E^2 - |\Delta|^2}]^{1/2}, & \sqrt{\mu^2+|\Delta|^2} < E, \\
\end{cases}
\end{align}
and the momentum of the hole-like left-moving excitation (green line in Fig.~\ref{fig:spectrum}) is
\begin{align}\label{eq:qhapp}
q_{h}(E) &= \frac{\sqrt{2m}}{\hbar} \begin{cases}
[\mu + \sqrt{E^2 - |\Delta|^2}]^{1/2}, & E < - \sqrt{\mu^2+|\Delta|^2}, \\
[\mu - i 0 + \sqrt{E^2 - |\Delta|^2}]^{1/2}, &  - \sqrt{\mu^2+|\Delta|^2} < E < - |\Delta|, \\
[\mu - i \sqrt{|\Delta|^2 - E^2}]^{1/2}, & -|\Delta| < E < |\Delta|, \\
[\mu - i 0 - \sqrt{E^2 - |\Delta|^2}]^{1/2}, & |\Delta| < E < \sqrt{\mu^2+|\Delta|^2}, \\
- i [\sqrt{E^2 - |\Delta|^2} - \mu]^{1/2}, & \sqrt{\mu^2+|\Delta|^2} < E. \\
\end{cases}
\end{align}
\end{widetext}
 The imaginary part  is chosen such that right-moving excitations with momenta $q_p(E)$ and $-q_h(E)$ decay at positive spatial infinity, and left-moving excitations with momenta $-q_p(E)$ and $q_h(E)$ with negative imaginary parts decay at negative spatial infinity. Note that the Andreev approximation neglects the energy-dependence of the momenta completely and assumes modes propagating with a fixed wave number \mbox{$k_F = \sqrt{2m\mu}/\hbar$}. The Bogoliubov coefficients for particle-like states are
\begin{subequations}\label{eq:uv}
\begin{align}
u_p(E) &= \begin{cases}
\sqrt{\dfrac{1}{2}\left(1 + \frac{\sqrt{E^2- |\Delta|^2}}{|E|}\right)}, & |E| \geq |\Delta|, \\
\sqrt{\dfrac{1}{2}\left(1+\frac{ i\sqrt{|\Delta|^2-E^2}}{E}\right)}, & |E| <|\Delta|,
\end{cases}  \\ 
v_p(E) &= \begin{cases}
\mathrm{sgn}(E)\sqrt{\dfrac{1}{2}\left(1-\frac{ \sqrt{E^2- |\Delta|^2}}{|E|}\right)}, & |E| \geq |\Delta|, \\
\sqrt{\dfrac{1}{2}\left(1-\frac{ i\sqrt{|\Delta|^2-E^2}}{E}\right)}, & |E| <|\Delta|.
\end{cases} 
\end{align}
\end{subequations}
Since $u_{p/h}^2 \equiv  \frac{1}{2} [1 + (\varepsilon_{q_{p/h}(E)}-\mu)/E]$ and $v_{p/h}^2 =  \frac{1}{2} [1 - (\varepsilon_{q_{p/h}(E)}-\mu)/E]$ where $q_p(E) = q_h^*(-E)$ and $q_p^2(E)/(2m) - \mu = -[q_h^2(E)/(2m) - \mu]$, we then have $u_p(E) = v_h^*(-E) = u_p^*(-E)$ and $v_p(E) = -u_h^*(-E) = -v_p^*(-E)$. For a given energy $E$, right- and left-moving particle eigenstates are equal; the same holds for right- and left-moving hole states. In addition, for a particle eigenstate $(u_p(E),v_p(E))^T$, the corresponding hole state of equal energy is $(u_h(E),v_h(E))^T = (v_p(E),u_p(E))^T$.

%++++++++++++++++++++++++++++++++++++++++
\begin{figure*}[t]
\includegraphics[width=0.8\linewidth]{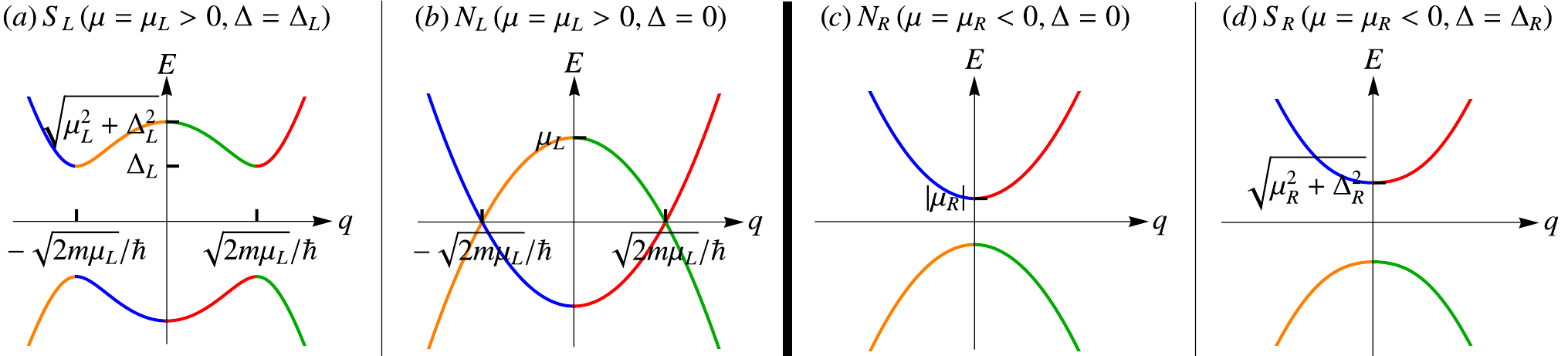}
\caption{
Example energy spectrum of the Bogoliubov excitations for the $S_L$-$N_L$-$I$-$N_R$-$S_R$ junction. Here, we choose $\mu_L >0$ and $\mu_R < 0$.
}
\label{fig:junction}
\end{figure*}
%++++++++++++++++++++++++++++++++++++++++

\section{Scattering states}\label{sec:scatteringstates}

In this appendix, we present the explicit form of the scattering states across the superfluid-normal-superfluid junction in terms of right- and left-moving Bogoliubov states in the superfluid (S$_L$ and S$_R$) and normal regions (N$_L$ and N$_R$). Figure~\ref{fig:junction} shows an example Bogoliubov spectrum across the superfluid-normal-superfluid junction, with the same color coding for the different branches as in Appendix~\ref{sec:bogoliubov}. We write the states with reference to the energy in the left reservoir, such that states in N$_R$ and S$_R$ carry an additional phase factor $e^{\mp iV\tau/\hbar}$. Since the Bogoliubov excitations are superpositions of a spin-up particle and a spin-down hole, a change in the reference potential (for example, due to the chemical potential mismatch between reservoirs) will affect the phase of the two Bogoliubov components in Eq.~\eqref{eq:eigenstate} in the opposite way~\cite{datta96}. While in the main text, we present only the calculation for the case where the normal-region length is $\ell \rightarrow 0$, for completeness, here we state the wave functions for junctions of finite length $\ell$.

The scattering state in the left superfluid region S$_L$ reads
\begin{widetext}
\begin{align}\label{eq:waveSL}
\Psi_{L}^{S,\zeta}(E) &= \sum_n e^{-iE_n\tau/\hbar} \Biggl\{
\delta_{n0} \delta_{\zeta,\rightarrow}\sqrt{\frac{m}{\hbar \qpL(E_n)}} \begin{pmatrix} \uL(E_n)\\\vL(E_n)\end{pmatrix} e^{i \qpL(E_n) (x+\ell/2)}\nonumber\\
 & + a_{1,n}^{\nub}  \sqrt{\frac{m}{\hbar \qpL(E_n)}} \begin{pmatrix}\uL(E_n)  \\\vL(E_n)\end{pmatrix} e^{i \qpL(E_n) (x+\ell/2)}+ b_{1,n}^{\nub} \sqrt{\frac{m}{\hbar \qhL(E_n)}} \begin{pmatrix}\vL(E_n)\\\uL(E_n)\end{pmatrix} e^{i \qhL(E_n) (x+\ell/2)} \nonumber\\
&+ c_{1,n}^{\nub}  \sqrt{\frac{m}{\hbar \qhL(E_n)}}\begin{pmatrix}\vL(E_n)\\\uL(E_n) \end{pmatrix} e^{- i \qhL(E_n) (x+\ell/2)}+ d_{1,n}^{\nub} \sqrt{\frac{m}{\hbar \qpL(E_n)}} \begin{pmatrix}\uL(E_n)\\\vL(E_n)\end{pmatrix} e^{- i \qpL(E_n) (x+\ell/2)}
\Biggr\} .
\end{align}
In the right superfluid region S$_R$, we have
\begin{align}\label{eq:waveSR}
&\Psi_{R}^{S,\zeta}(E) = \sum_n e^{-iE_n\tau/\hbar}\Biggl\{\delta_{n0} \delta_{\zeta,\leftarrow}\sqrt{\frac{m}{\hbar \qpR(E_n)}} \begin{pmatrix} \uR(E_n) e^{i\phi - iV\tau/\hbar}\\\vR(E_n)e^{iV\tau/\hbar}\end{pmatrix} e^{-i \qpR(E_n) (x-\ell/2)}\nonumber\\
&+a_{2,n}^{\nub}  \sqrt{\frac{m}{\hbar \qpR(E_n)}} \begin{pmatrix}\uR(E_n) e^{i\phi - iV\tau/\hbar} \\\vR(E_n) e^{iV\tau/\hbar}\end{pmatrix} e^{i \qpR(E_n) (x-\ell/2)}+ b_{2,n}^{\nub} \sqrt{\frac{m}{\hbar \qhR(E_n)}} \begin{pmatrix}\vR(E_n)e^{i\phi - iV\tau/\hbar}\\\uR(E_n)e^{iV\tau/\hbar}\end{pmatrix} e^{i \qhR(E_n) (x-\ell/2)}\nonumber\\
&+ c_{2,n}^{\nub}  \sqrt{\frac{m}{\hbar \qhR(E_n)}}\begin{pmatrix}\vR(E_n) e^{i\phi - iV\tau/\hbar}\\\uR(E_n) e^{iV\tau/\hbar}\end{pmatrix} e^{- i \qhR(E_n) (x-\ell/2)}+ d_{2,n}^{\nub} \sqrt{\frac{m}{\hbar \qpR(E_n)}} \begin{pmatrix}\uR(E_n)e^{i\phi - iV\tau/\hbar}\\\vR(E_n) e^{iV\tau/\hbar}\end{pmatrix} e^{- i \qpR(E_n) (x-\ell/2)}
\Biggr\} ,
\end{align}
where the phase factors $e^{\mp iV\tau/\hbar}$ account for the chemical potential mismatch between the reservoirs. Likewise, the wave function in the left normal region N$_L$ is
\begin{align}
\Psi_{L}^{N,\zeta}(E) &= \sum_n e^{-iE_n\tau/\hbar}\Bigl\{
a_{L,n}^{\nub} \sqrt{\frac{m}{\hbar k_{pL}(E_n)}} \begin{pmatrix}1\\0\end{pmatrix} e^{i k_{pL}(E_n) (x+\ell/2)}+ d_{L,n}^{\nub} \sqrt{\frac{m}{\hbar k_{pL}(E_n)}} \begin{pmatrix}1\\0\end{pmatrix} e^{- i k_{pL}(E_n) (x+\ell/2)}
\Bigr\} \nonumber \\*
&+ \sum_n e^{-i E_n\tau/\hbar} \Bigl\{
b_{L,n}^{\nub} \sqrt{\frac{m}{\hbar k_{hL}(E_n)}} \begin{pmatrix}0\\1\end{pmatrix} e^{i k_{hL}(E_n) (x+\ell/2)} + c_{L,n}^{\nub} \sqrt{\frac{m}{\hbar k_{hL}(E_n)}} \begin{pmatrix}0\\1\end{pmatrix} e^{- i k_{hL}(E_n) (x+\ell/2)}
\Bigr\}, \label{eq:waveNL}
\end{align}
and in the right normal region N$_R$, we have
\begin{align}\label{eq:waveNR}
&\Psi_{R}^{N,\zeta}(E) = \sum_n e^{-i(E_n-V)\tau/\hbar} \Bigl\{
a_{R,n}^{\nub} \sqrt{\frac{m}{\hbar \kpR(E_n)}} \begin{pmatrix}1\\0\end{pmatrix} e^{i \kpR(E_n) (x-\ell/2)}
+ d_{R,n}^{\nub}\sqrt{\frac{m}{\hbar \kpR(E_n)}} \begin{pmatrix}1\\0\end{pmatrix} e^{- i \kpR(E_n) (x-\ell/2)}
\Bigr\} \nonumber \\*
&+ \sum_n e^{-i(E_n+V)\tau/\hbar} \Bigl\{
b_{R,n}^{\nub} \sqrt{\frac{m}{\hbar \khR(E_n)}} \begin{pmatrix}0\\1\end{pmatrix} e^{i \khR(E_n) (x-\ell/2)}+ c_{R,n}^{\nub} \sqrt{\frac{m}{\hbar \khR(E_n)}} \begin{pmatrix}0\\1\end{pmatrix} e^{- i \khR(E_n) (x-\ell/2)}
\Bigr\} ,
\end{align}
\end{widetext}
with  momenta
\begin{subequations}\label{eq:kpapp} 
\begin{align}
 k_{p,L/R}(E) &= \frac{\sqrt{2m}}{\hbar}  \sqrt{E + \mu_{L/R} + i0}, \\
 k_{h,L/R}(E) &= \frac{\sqrt{2m}}{\hbar}  \sqrt{- E + \mu_{L/R} - i0} ,
 \end{align}
\end{subequations}
where the chemical potential of the left (N$_L$) and right (N$_R$) normal region are taken to be $\mu_{L}$ and $\mu_R$ which are the same as the chemical potential of the left (S$_L$) and right (S$_R$) reservoir, respectively. The coefficients in the normal region are shown in Fig.~\ref{fig:1} of the main text.

In terms of the scattering state in the normal region, Eq.~\eqref{eq:waveNL}, the expression [Eq.~\eqref{eq:Tphm}] for the dimensionless current density becomes (we evaluate the current in the left normal region)
\begin{widetext}
\begin{subequations}\label{eq:Tphzeta}
\begin{align}
T_{p,\zeta}^{p(l)}(E) &= \frac{k_{pL}(E_n) + k_{pL}^*(E_{n+l})}{2 \sqrt{k_{pL}(E_n) k_{pL}^*(E_{n+l})}} \biggl[(a_{L,n+l}^{\nub})^* a_{L,n}^{\nub} - (d_{L,n+l}^{\nub})^* d_{L,n}^{\nub}\biggr] \nonumber\\*
&\qquad+ \frac{k_{pL}(E_n) - k_{pL}^*(E_{n+l})}{2 \sqrt{k_{pL}(E_n) k_{pL}^*(E_{n+l})}} \biggl[(d_{L,n+l}^{\nub})^* a_{L,n}^{\nub} - (a_{L,n+l}^{\nub})^* d_{L,n}^{\nub}\biggr], \\
T_{p,\zeta}^{h(l)}(E) &= \frac{k_{hL}(E_n) + k_{hL}^*(E_{n+l})}{2 \sqrt{k_{hL}(E_n) k_{hL}^*(E_{n+l})}} \biggl[(b_{L,n+l}^{\nub})^* b_{L,n}^{\nub} - (c_{L,n+l}^{\nub})^* c_{L,n}^{\nub}\biggr] \nonumber\\*
&\qquad+ \frac{k_{hL}(E_n) - k_{hL}^*(E_{n+l})}{2 \sqrt{k_{hL}(E_n) k_{hL}^*(E_{n+l})}} \biggl[(c_{L,n+l}^{\nub})^* b_{L,n}^{\nub} - (b_{L,n+l}^{\nub})^* c_{L,n}^{\nub}\biggr].
\end{align}
\end{subequations}
\end{widetext}
For a dc-current, which corresponds to $l = 0$, Eq.~\eqref{eq:Tphzeta} reduces to Eq.~\eqref{eq:Tpzeta} of the main text. The challenge is to determine the scattering coefficients $\{a_{L,n}^\zeta\}$, $\{b_{L,n}^\zeta\}$, $\{c_{L,n}^\zeta\}$, and $\{d_{L,n}^\zeta\}$ from a solution of the scattering problem, which is done in the next Appendix~\ref{sec:matching}. 

Finally, using $e^{- i q_h(E) x} = \bigl[e^{i q_p(-E) x}\bigr]^*$ and $(u(E), v(E)) = (v(-E), -u(-E))^*$, it follows that that the current due to quasihole injections is equal to the current due to quasiparticle injections. As a result, the total current is equal to twice the current due to quasiparticle injections, which justifies the factor of $2$ in Eq.~\eqref{eq:currentfourier}.

\section{Scattering coefficients and wave function matching}\label{sec:matching}

The scattering states states introduced in Sec.~\ref{sec:scatteringmatrix} and Appendix~\ref{sec:scatteringstates} contain a large number of scattering amplitudes $\{a_{j,n}^\zeta\}$, $\{b_{j,n}^\zeta\}$, $\{c_{j,n}^\zeta\}$, $\{d_{j,n}^\zeta\}$ in each region $j=1,L,R,2$. As 
discussed in Sec.~\ref{sec:scatteringmatrix}, these amplitudes are linked by various individual scattering processes, which are determined by a solution of the scattering problem in the potential given by Eqs.~\eqref{eq:mupotential} and~\eqref{eq:potential}. They are obtained by matching the scattering wave functions and their derivatives at the normal-superfluid boundaries and at the tunnel barrier. Here, we present the analytic expressions of the coefficients in the scattering matrix at the left superfluid-normal boundary [Eq.~\eqref{eq:resmatchingL}]:

\begin{widetext}
\begin{subequations}\label{eq:scattcoeff}
\begin{align}
\NLnp &= \frac{
\uL^2(E_n) [k_{hL}(E_n)+\qhL(E_n)] [k_{pL}(E_n)-\qpL(E_n))] - 
\vL^2(E_n) [\qhL(E_n)+k_{pL}(E_n)] [k_{hL}(E_n)-\qpL(E_n)]
}{
\uL^2(E_n) [k_{hL}(E_n)+\qhL(E_n)] [k_{pL}(E_n)+\qpL(E_n)] + 
\vL^2(E_n) [\qhL(E_n)-k_{pL}(E_n)] [k_{hL}(E_n)-\qpL(E_n)]
} \nonumber \\*
&\qquad \stackrel{ AA}{\to} 0, \label{eq:NLnp} \\ 
\NLnh &= \frac{
\uL^2(E_n) [k_{hL}(E_n)-\qhL(E_n)] [k_{pL}(E_n)+\qpL(E_n)] + 
\vL^2(E_n) [\qhL(E_n)-k_{pL}(E_n)] [k_{hL}(E_n)+\qpL(E_n)]
}{
\uL^2(E_n) [k_{hL}(E_n)+\qhL(E_n)] [k_{pL}(E_n)+\qpL(E_n)] + 
\vL^2(E_n) [\qhL(E_n)-k_{pL}(E_n)] [k_{hL}(E_n)-\qpL(E_n)]
} \nonumber \\*
&\qquad \stackrel{ AA}{\to} 0 , \label{eq:NLnh} \\ 
\ALn &= \frac{2 \sqrt{k_{pL}(E_n) k_{hL}(E_n)} \uL(E_n) \vL(E_n) (\qhL(E_n)+\qpL(E_n))}{\uL^2(E_n) [k_{hL}(E_n)+\qhL(E_n)] [k_{pL}(E_n)+\qpL(E_n)] + \vL^2(E_n) [\qhL(E_n)-k_{pL}(E_n)] [k_{hL}(E_n)-\qpL(E_n)]} \nonumber \\*
&  \qquad \stackrel{ AA}{\to} \frac{\vL(E_n)}{\uL(E_n)} , \label{eq:ALn} 
\end{align}
\end{subequations}
and
\begin{subequations}\label{eq:JLn}
\begin{align}
\JLnp &= \frac{2 \sqrt{k_{pL}(E_n) \qpL(E_n)} [\qhL(E_n) + k_{hL}(E_n)] \uL(E_n) \left(\uL^2(E_n)-\vL^2(E_n)\right)}{\uL^2(E_n) [k_{hL}(E_n)+\qhL(E_n)] [k_{pL}(E_n)+\qpL(E_n)] + \vL^2(E_n) [\qhL(E_n)-k_{pL}(E_n)] [k_{hL}(E_n)-\qpL(E_n)]} \nonumber \\
&  \qquad\stackrel{ AA}{\to} \frac{\uL^2(E_n)-\vL^2(E_n)}{\uL(E_n)} , \label{eq:JLnp} \\
\JLnh&= \frac{2 \sqrt{k_{hL}(E_n) \qpL(E_n)} [\qhL(E_n) - k_{pL}(E_n)] \vL(E_n) \left(\uL^2(E_n)-\vL^2(E_n)\right)}{\uL^2(E_n) [k_{hL}(E_n)+\qhL(E_n)][k_{pL}(E_n)+\qpL(E_n)] + \vL^2(E_n) [\qhL(E_n)-k_{pL}(E_n)] [k_{hL}(E_n)-\qpL(E_n)]} \nonumber \\
&  \qquad \stackrel{ AA}{\to} 0 \label{eq:JLnh} .
\end{align}
\end{subequations}
\end{widetext}
The scattering coefficients in Eqs.~\eqref{eq:scattcoeff} and~\eqref{eq:JLn} are plotted in Fig.~\ref{fig:S4} of the main text. In the above equations, we also include the limiting form of the scattering coefficients when using the Andreev approximation (AA). Results in the Andreev approximation limit agree with Refs.~\cite{blonder82,beenakker92}. Taking into account that there are no particle (hole) propagating states for $E_n < -\mu_L$ ($E_n>\mu_L$), we have the scattering coefficients given by
\begin{subequations}\label{eq:S4}
\begin{align}
\NLnp &=  \begin{cases}
\mathrm{Eq.}~\eqref{eq:NLnp}, &  E_n \geq -\mu_L,\\
 1, & E_n < -\mu_L,
\end{cases}\\
\NLnh &=  \begin{cases}
\mathrm{Eq.}~\eqref{eq:NLnh}, & E_n \leq \mu_L,  \\
1, & E_n  > \mu_L, 
\end{cases}\\
\ALn &=  \begin{cases}
\mathrm{Eq.}~\eqref{eq:ALn}, & -\mu_L\leq E_n \leq \mu_L  \,\,\mathrm{for}\,\, \mu_L \geq 0, \\
0, &  \mathrm{otherwise.}
\end{cases}
\end{align}
\end{subequations}

\section{Recurrence relation for the scattering amplitudes}\label{sec:recur}

In this Appendix, we derive a recurrence relation for the scattering amplitudes in the normal region, which are needed to compute the current density Eq.~\eqref{eq:Tpzeta} across the junction. The starting point is the following set of equations that contains only the coefficients of normal-region modes propagating away from the tunnel barrier, which are obtained by expressing the incoming modes in the scattering matrices [Eq.~\eqref{eq:scatbarrier}] in terms of states reflected at the normal-superfluid boundaries or transmitted from the reservoir [Eqs.~\eqref{eq:resmatchingL} and~\eqref{eq:resmatchingR}]: 

\begin{widetext}
\begin{subequations}
\begin{align}
d_{L,n}^{\rightarrow} 
&=
r_{p,n} 					\Bigl[ \NLnp  d_{L,n}^{\rightarrow} + \ALn b_{L,n}^{\rightarrow} + \delta_{n0} \JLnp \Bigr]+
t_{p,n} 	\Bigl[ \NRnpp a_{R,n+1}^{\rightarrow} + e^{i\phi} \ARnp c_{R,n+1}^{\rightarrow}\Bigr], 
\label{eq:cond1} \\
a_{R,n+1}^{\rightarrow} 
&= 
t_{p,n} 					\Bigl[\NLnp  d_{L,n}^{\rightarrow} + \ALn b_{L,n}^{\rightarrow} + \delta_{n0} \JLnp \Bigr] 
-
\frac{t_{p,n}}{t_{p,n}^*} r_{p,n}^*\Bigl[ \NRnpp  a_{R,n+1}^{\rightarrow} + e^{i\phi} \ARnp  c_{R,n+1}^{\rightarrow}\Bigr], 
\label{eq:cond2} \\
b_{L,n}^{\rightarrow} 
&= 
r_{h,n} 					\Bigl[ \ALn d_{L,n}^{\rightarrow} + \NLnh  b_{L,n}^{\rightarrow} + \delta_{n0} \JLnh\Bigr] +
t_{h,n} 	\Bigl[e^{-i\phi} \ARnm  a_{R,n-1}^{\rightarrow} + \NRnhm c_{R,n-1}^{\rightarrow}\Bigr], 
\label{eq:cond3} \\
c_{R,n-1} ^{\rightarrow}
&= 
t_{h,n} 					\Bigl[\ALn  d_{L,n}^{\rightarrow} + \NLnh b_{L,n}^{\rightarrow} + \delta_{n0}\JLnh \Bigr]-
\frac{t_{h,n}}{t_{h,n}^*} r_{h,n}^* \Bigl[e^{-i\phi} \ARnm a_{R,n-1}^{\rightarrow} + \NRnhm  c_{R,n-1}^{\rightarrow}\Bigr] 
\label{eq:cond4} .
\end{align}
\end{subequations}
\end{widetext}
Next, we solve the last two equations [Eqs.~\eqref{eq:cond3} and~\eqref{eq:cond4}] for the coefficients $\{a_{R,n}^{\rightarrow}\}$ and $\{c_{R,n}^{\rightarrow}\}$ of the right-normal region and 
substitute the result for $a_{R,n-1}^{\rightarrow}$ and $c_{R,n-1}^{\rightarrow}$ in the remaining two constraint equations [Eqs.~\eqref{eq:cond1} and~\eqref{eq:cond2}], which only leaves the coefficients $\{b_{L,n}^{\rightarrow}\}$ and $\{d_{L,n}^{\rightarrow}\}$ as unknown variables. Once these are known, the remaining coefficients $a_{L,n}^{\rightarrow}$ and $c_{L,n}^{\rightarrow}$ in the left normal region follow from Eq.~\eqref{eq:resmatchingL}. We solve the constraint equations~[Eqs.~\eqref{eq:cond1} and~\eqref{eq:cond2}] for $b_{L,n}^{\rightarrow}$ and express the $\{b_{L,n}^{\rightarrow}\}$ in terms of the $d_{L,n}^{\rightarrow}$ coefficients as
\begin{widetext}
\begin{align}
&b_{L,n+2}^{\rightarrow} = 
\frac{z_{n+1}}{r_{h,n+2}^*} \biggl\{
|t_{p,n}|^2 \ARnp d_{L,n}^{\rightarrow} |t_{h,n+2}|^2\nonumber \\*
&+ \biggl[t_{h,n+2} \Bigl(t_{p,n} \left( \NRnhp \NRnpp - \ARnp^2\Bigr)+r_{p,n}
   t_{p,n}^* \NRnhp \right)+r_{h,n+2} t_{h,n+2}^* 
	\left(t_{p,n}
   \NRnpp +r_{p,n} t_{p,n}^*\right)\biggr] \frac{ \ALnpt }{|t_{h,n+2}|^2} d_{L,n+2}^{\rightarrow} \nonumber \\*
&+ \biggl[t_{h,n+2} \Bigl(t_{p,n} \left( \NRnhp \NRnpp -\ARnp^2\Bigr)+r_{p,n}
   t_{p,n}^* \NRnhp \right)+r_{h,n+2} t_{h,n+2}^* \left(t_{p,n}
   \NRnpp +r_{p,n} t_{p,n}^*\right)\biggr] \frac{\JLnhpt }{|t_{h,n+2}|^2} \delta _{n+2,0} 
   \biggr\},
   \label{eq:getb}
\end{align}
with
\begin{align}\label{eq:zn}
z_{n}&= r_{h,n+1}^* |t_{h,n+1}|^2 \biggl\{
t_{h,n+1} (\NLnhp  - r_{h,n+1}^*)\Bigl[ t_{p,n-1} (\ARn^2 - \NRnh \NRnp) - r_{p,n-1} t_{p,n-1}^* \NRnh\Bigr]\nonumber \\*
&\qquad + 
t_{h,n+1}^* (1 - r_{h,n+1} \NLnhp ) (t_{p,n-1} \NRnp + r_{p,n-1} t_{p,n-1}^*)
\biggr\}^{-1}.
\end{align}
Finally, using this result in the remaining constraint equation [Eq.~\eqref{eq:cond1}], we obtain a closed-form matrix equation for the $\{d_{L,n}\}$, which must be solved first to determine all other scattering amplitudes [Eq.~\eqref{eq:recurrence} of the main text]
\begin{align}
\alpha_n^{\rightarrow} d_{L,n+2}^{\rightarrow} + \beta_n^{\rightarrow} d_{L,n}^{\rightarrow} + \gamma_n^{\rightarrow} d_{L,n-2}^{\rightarrow} &= \SLnpl  \delta_{n0} + \SLnhl \delta_{n+2,0}, \label{eq:getd} 
\end{align}
with coefficients
\begin{subequations}\label{eq:coeffrecur}
\begin{align}
%\alpha_n
\alpha_n^{\rightarrow} &= - \frac{r_{p,n}}{r_{h,n+2}^*} |t_{p,n}|^2 \ALnpt \ARnp z_{n+1} ,
\\[1ex]
%\beta_n
\beta_n^{\rightarrow} &= 1-r_{p,n} \NLnp + \frac{r_{p,n}}{r_{h,n}^*} \ALn^2 \Bigl[t_{p,n-2}
    \Bigl(t_{h,n} \left( \ARnm^2- \NRnhm \NRnpm \right)-    t_{h,n}^* r_{h,n} 
    \NRnpm \Bigr)  \nonumber \\
   &
	\qquad\qquad - t_{p,n-2}^* r_{p,n-2} \Bigl(t_{h,n} \NRnhm +t_{h,n}^*r_{h,n}
    \Bigr)\Bigr] \frac{z_{n-1}}{|t_{h,n}|^2} - \frac{t_{p,n}^*}{r_{h,n+2}^*} t_{p,n}^2 \Bigl[
    t_{h,n+2}^* \NRnpp  \left(1-r_{h,n+2}
    \NLnhpt \right)
    \nonumber \\
&\qquad\qquad +t_{h,n+2} \left( \NLnhpt -r_{h,n+2}^*\right)\left(\ARnp^2- \NRnhp \NRnpp \right) \Bigr] \frac{z_{n+1}}{|t_{h,n+2}|^2}, \\[1ex]
%\gamma_n
\gamma_n^{\rightarrow} &= - \frac{r_{p,n}}{r_{h,n}^*} |t_{p,n-2}|^2 \ALn \ARnm  z_{n-1} , \\[1ex]
%S^p_{1,n}
\SLnpl &= r_{p,n} \JLnp - \frac{r_{p,n}}{r^*_{h,n}} \JLnh  \ALn  \Bigr[
t_{p,n-2} \Bigl(t_{h,n} 
( \ARnm^2 - \NRnhm   \NRnpm )- t_{h,n}^* r_{h,n}
\NRnpm \Bigr) \nonumber \\
&\qquad\qquad - t_{p,n-2}^* r_{p,n-2}
\Bigl(t_{h,n} \NRnhm  + t_{h,n}^* r_{h,n}\Bigr)
\Bigr] \frac{z_{n-1}}{|t_{h,n}|^2}, \\[1ex]
%S^h
\SLnhl &= \frac{r_{p,n}}{r_{h,n+2}^*} |t_{p,n}|^2 \JLnhpt  \ARnp  z_{n+1} .
\end{align}
\end{subequations}
\end{widetext}

For quasiparticles injected from the right reservoir ($\zeta = \leftarrow$), we derive a similar recurrence relation
\begin{align}\label{eq:recura}
\alpha_n^{\leftarrow} a_{R,n-2}^{\leftarrow} + \beta_n^{\leftarrow} a_{R,n}^{\leftarrow} + \gamma_n^{\leftarrow} a_{R,n+2}^{\leftarrow} &= \SRnpr  \delta_{n0} + \SRnhr \delta_{n-2,0} , 
\end{align}
where the coefficients $\alpha_n^{\leftarrow}$,  $\beta_n^{\leftarrow}$, $\gamma_n^{\leftarrow}$, $\SRnpr$, and $\SRnhr$ are obtained from the corresponding quantities in Eq.~\eqref{eq:coeffrecur} by replacing $L \rightarrow R$, $n\pm 1 \rightarrow n \mp 1$, and $n\pm 2 \rightarrow n \mp 2$. \\

As a check of our results, consider the Andreev approximation: Here, Eqs.~\eqref{eq:zn} and~\eqref{eq:coeffrecur} simplify considerably and reduce to
\begin{align}\label{eq:znAA}
z_{n}= \biggl[\frac{r_{p,n-1} t_{p,n-1}^*}{t_{h,n+1}r_{h,n+1}^*} - \frac{t_{p,n-1}}{t_{h,n+1}^*}A_{R,n}^2 \biggr]^{-1},
\end{align}
and 
\begin{subequations}\label{eq:coeffrecurAA}
\begin{align}
\alpha_n^{\rightarrow} &= - \frac{r_{p,n}}{r_{h,n+2}^*} |t_{p,n}|^2 \ALnpt \ARnp z_{n+1},
\\
\beta_n^{\rightarrow} &= 1- \frac{r_{p,n}}{r_{h,n}^*} \ALn^2 z_{n-1}\Bigl[\frac{r_{p,n-2}r_{h,n}t_{p,n-2}^*}{t_{h,n}} \nonumber\\
& \quad -\frac{t_{p,n-2}}{t_{h,n}^*}\ARnm^2\Bigr]+ \frac{t_{p,n}}{t_{h,n+2}^*} |t_{p,n}|^2\ARnp^2 z_{n+1}, \\
\gamma_n^{\rightarrow} &= - \frac{r_{p,n}}{r_{h,n}^*} |t_{p,n-2}|^2 \ALn \ARnm  z_{n-1}, \\
\SLnpl &= r_{p,n} \JLnp,  \\
\SLnhl &= 0,
\end{align}
\end{subequations}
respectively, which agrees with the literature~\cite{hurd97}. 

\section{Solution of the recurrence relations for the scattering amplitudes}\label{sec:Lenz}

In this section, we discuss the solution of the infinite-dimensional matrix equation [Eq.~\eqref{eq:recurrence}]. The solution can be obtained by first casting a continued-fraction representation of the constraint equations (see Refs.~\cite{bratus95,averin95,hurd97}), which is then solved using a modified Lentz method~\cite{press02}. Other scattering amplitudes then follow by direct substitution. 

We begin by defining
\begin{align}\label{eq:defxn}
x_n^{\rightarrow} 
= 
\begin{cases}
\dfrac{d_{L,n}^{\rightarrow}}{d_{L,n-2}^{\rightarrow}}, & n>2, \\[3ex]
\dfrac{d_{L,n}^{\rightarrow}}{d_{L,n+2}^{\rightarrow}}, & n<0,
\end{cases} 
\end{align}
and rewriting Eq.~\eqref{eq:getd} as
\begin{subequations}
\begin{align}
\alpha_n^{\rightarrow} x_{n+2}^{\rightarrow} + \beta_n^{\rightarrow} + \frac{\gamma_n^{\rightarrow}}{x_n^{\rightarrow}} &= 0, \qquad n>0, \label{eq:cont1} \\
\frac{\alpha_n^{\rightarrow}}{x_n^{\rightarrow}} + \beta_n^{\rightarrow} + \gamma_n^{\rightarrow} x_{n-2}^{\rightarrow} &= 0, \qquad n<-2, \label{eq:cont2}
\end{align}
\end{subequations}
with two additional equations containing the source terms for $n=0$ and for $n=-2$, i.e.,
\begin{subequations}\label{eq:nzerotwob}
\begin{align}
(\beta_0^{\rightarrow} + \alpha_0^{\rightarrow} x_{2}^{\rightarrow}) d_{L,0}^{\rightarrow} + \gamma_0^{\rightarrow} d_{L,-2}^{\rightarrow} &= \SLzp, \\
\alpha_{-2}^{\rightarrow} d_{L,0}^{\rightarrow} + (\beta_{-2}^{\rightarrow} + \gamma_{-2}^{\rightarrow} x_{-4}^{\rightarrow}) d_{L,-2}^{\rightarrow} &= \SLnth.
\end{align}
\end{subequations}
We rewrite Eq.~\eqref{eq:nzerotwob} in matrix form as
\begin{widetext}
\begin{align}
\begin{pmatrix}d_{L,0}^{\rightarrow}\\ d_{L,-2}^{\rightarrow}\end{pmatrix}
&=
\frac{1}{(\beta_0^{\rightarrow} + \alpha_0 x_2^{\rightarrow}) (\beta_{-2}^{\rightarrow} + \gamma_{-2}^{\rightarrow} x_{-4}^{\rightarrow}) - \alpha_{-2}^{\rightarrow} \gamma_0^{\rightarrow}}
\begin{pmatrix}
\beta_{-2}^{\rightarrow} + \gamma_{-2}^{\rightarrow} x_{-4}^{\rightarrow} & - \gamma_0^{\rightarrow} \\ 
- \alpha_{-2}^{\rightarrow} & \beta_0^{\rightarrow} + \alpha_0^{\rightarrow} x_{2}^{\rightarrow}
\end{pmatrix}
\begin{pmatrix} \SLzp\\ \SLnth \end{pmatrix} ,
\end{align}
\end{widetext}
where the coefficients $d_{L,0}^{\rightarrow}$ and $d_{L,-2}^{\rightarrow}$ can be calculated once the $\{x_n^{\rightarrow}\}$ are known. Having solved for $d_{L,0}^{\rightarrow}$ and $d_{L,-2}^{\rightarrow}$, we obtain the remaining coefficients $\{d_{L,n}^{\rightarrow}\}_{n\neq 0,-2}$ from Eq.~\eqref{eq:defxn},
\begin{align}
d_{L,n}^{\rightarrow} &= \begin{cases}
x_n^{\rightarrow} x_{n-2}^{\rightarrow} \ldots x_2^{\rightarrow} d_{L,0}^{\rightarrow}, & n>0,\\[1ex]
x_n^{\rightarrow}x_{n+2}^{\rightarrow} \ldots x_{-4}^{\rightarrow} d_{L,-2}^{\rightarrow}, & n<-2.
\end{cases} 
\end{align}

To determine the values of $\{x_n^{\rightarrow}\}$, we rewrite Eqs.~\eqref{eq:cont1} and~\eqref{eq:cont2} as continued fraction expansions, where we express $x_n^{\rightarrow}$ in terms of higher-index coefficients $x_m^{\rightarrow}$ with $|m| = |n|+2$:
\begin{align}
x_n^{\rightarrow}  =\begin{cases} \dfrac{-\gamma_n^{\rightarrow}}{\beta_n^{\rightarrow} + \alpha_n^{\rightarrow} x_{n+2}^{\rightarrow}}, &n>0, \label{eq:fraction1} \\[2ex]
 \dfrac{-\alpha_n^{\rightarrow}}{\beta_n^{\rightarrow} + \gamma_n^{\rightarrow} x_{n-2}^{\rightarrow}}, & n<-2. 
\end{cases} 
\end{align}
Recasting this in the general form of a continued fraction gives
\begin{align}
x_n^{\rightarrow} 
&= 
g_{n,0}^{\rightarrow} + \cfrac{f_{n,1}^{\rightarrow}}{g_{n,1}^{\rightarrow} \, + \, \cfrac{f_{n,2}^{\rightarrow}}{g_{n,2}^{\rightarrow} \, + \, \cfrac{f_{n,3}^{\rightarrow}}{g_{n,3}^{\rightarrow} \, + \cdots \,}}}, \label{eq:continuedfraction}
\end{align}
with the following coefficients for $n>0$:
\begin{align}
f_{n,m}^{\rightarrow} 
&= 
\begin{cases} 
- \gamma_n^{\rightarrow}, & m=1, \\[1ex] 
- \alpha_{n+2(m-2)}^{\rightarrow} \gamma_{n+2(m-1)}^{\rightarrow}, & m> 1,
\end{cases} 
\\
g_{n,m}^{\rightarrow} 
&= 
\begin{cases} 
0, & m=0, \\[1ex] 
\beta_{n+2(m-1)}^{\rightarrow}, & m> 0,
\end{cases}
\end{align}
and for $n<-2$:
\begin{align}
f_{n,m}^{\rightarrow} 
&= 
\begin{cases} 
- \alpha_n^{\rightarrow}, & m=1, \\[1ex] 
- \gamma_{n-2(m-2)}^{\rightarrow} \alpha_{n-2(m-1)}^{\rightarrow}, & m> 1,
\end{cases} \\
g_{n,m}^{\rightarrow} 
&= 
\begin{cases} 
0, & m=0, \\[1ex] 
\beta_{n-2(m-1)}^{\rightarrow}, & m> 0.
\end{cases} 
\end{align}

In principle, the continued fraction expansion [Eq.~\eqref{eq:continuedfraction}] can be solved by introducing an index $n_{\rm max}$ with $x^\rightarrow_{n_{\rm max}} = 0$ and iterating lower-index coefficients $x_n^\rightarrow$, which however introduces a systematic error. Instead, we employ the modified Lentz method~\cite{press02} to solve continued fraction systems [Eq.~\eqref{eq:continuedfraction}], which we outline in the remainder of the section (restricting to the case $n>0$ for notational simplicity). First, we define $x_n^{\rightarrow,(m)}$ (with $m>0$) as the partial evaluation of the coefficient $x_n^{\rightarrow}$ obtained by setting $x_{n+m}^{\rightarrow}=0$ in Eq.~\eqref{eq:continuedfraction}. Formally, this is written as
\begin{align}
x_n^{\rightarrow,(m)} &= \frac{F_m^{\rightarrow}}{G_m^{\rightarrow}}, 
\end{align}
with $F_m^{\rightarrow} = g_m^{\rightarrow} F_{m-1}^{\rightarrow} + f_m^{\rightarrow} F_{m-2}^{\rightarrow}$ and $G_m^{\rightarrow} = g_m^{\rightarrow} G_{m-1}^{\rightarrow} + f_m^{\rightarrow} G_{m-2}^{\rightarrow}$, where $F_0^{\rightarrow}=g_0^{\rightarrow}$, $F_{-1}^{\rightarrow}=1$, $G_0^{\rightarrow}=1$, and $G_{-1}^{\rightarrow}=0$ as well as $f_1^{\rightarrow}=-\alpha_n^{\rightarrow}$, $g_{n\geq2}^{\rightarrow} = - \gamma_{n+2(m-1)}^{\rightarrow} \alpha_{n+2m}^{\rightarrow}$ and $g_0^{\rightarrow}=0$, $g_{n\geq1}^{\rightarrow} = \beta_{n+2(m-1)}^{\rightarrow}$. The modified Lentz method introduces the ratios $W_m^{\rightarrow}=F_m^{\rightarrow}/F_{m-1}^{\rightarrow}$ and $Y_m^{\rightarrow} = G_{m-1}^{\rightarrow}/G_m^{\rightarrow}$, such that $x_n^{\rightarrow,(m)} = x_n^{\rightarrow,(m-1)} W_m^{\rightarrow} Y_m^{\rightarrow}$, and then iterates with initial conditions $W_0^{\rightarrow} = x_n^{\rightarrow,(0)}$ and $Y_0^{\rightarrow} = 0$ as follows:
\begin{align}
Y_m^{\rightarrow} &= \frac{1}{g_m^{\rightarrow} + f_m^{\rightarrow} Y_{m-1}^{\rightarrow}}, \\
W_m^{\rightarrow} &= g_m^{\rightarrow}+ \frac{f_m^{\rightarrow}}{W_{m-1}^{\rightarrow}} .
\end{align}
This is done until convergence is reached, i.e., $x_n^{\rightarrow,(m)}$ does not change within the numerical resolution. Whenever $W_m^{\rightarrow}$ or $(Y_{m}^{\rightarrow})^{-1}$ are zero, they should be shifted by an infinitesimal amount. Similarly, by using the substitution $d_{L,n}^{\rightarrow} \rightarrow a_{R,n}^{\leftarrow}$ and $d_{L,n\pm 2}^{\rightarrow} \rightarrow a_{R,n\mp 2}^{\leftarrow}$, we use the above method to solve the recurrence relation [Eq.~\eqref{eq:recura}] corresponding to the case where quasiparticles are injected from the right superfluid.

\section{Analytical expression for the tunneling limit}\label{app:tunneling}

In this appendix, we derive an analytic expression for the current in the tunneling limit ($\mathcal{T} = t^2 \ll 1$) and show that it takes the form given in Eq.~\eqref{eq:tunnelling}. Substituting the perturbative transmission and reflection parameters into Eqs.~\eqref{eq:resmatchingL}-\eqref{eq:scatbarrier}, we find the following relations between the scattering amplitudes (see Fig.~\ref{fig:scatteringstates} for a graphical illustration of the various coefficients) for the case of quasiparticles injected from the left reservoir:
\begin{widetext}
\begin{alignat}{2}\label{eq:tunnelingrel}
a_{L,2}^{\rarr} &= N_{L,2}^p d_{L,2}^{\rarr} + A_{L,2} b_{L,2}^{\rarr}, \nonumber\\
b_{L,2}^{\rarr} &= -c_{L,2}^{\rarr}, \nonumber\\
c_{L,2}^{\rarr} &= N_{L,2}^h b_{L,2}^{\rarr} + A_{L,2} d_{L,2}^{\rarr},\nonumber \\
d_{L,2}^{\rarr} &= - a_{L,2}^{\rarr} ,\nonumber \\[-6ex]
& & \qquad \qquad a_{R,1}^{\rarr} &= -i t a_{L,0}^{\rarr} - d_{R,1}^{\rarr}, \nonumber \\
& & \qquad \qquad b_{R,1}^{\rarr} &= A_{R,1} a_{R,1}^{\rarr} + N_{R,1}^h c_{R,1}^{\rarr}, \nonumber \\
& & \qquad \qquad c_{R,1}^{\rarr} &= - b_{R,1}^{\rarr}, \nonumber \\
& & \qquad \qquad d_{R,1}^{\rarr} &= N_{R,1}^p a_{R,1}^{\rarr} + A_{R,1} c_{R,1}^{\rarr}, \nonumber \\[-6ex]
a_{L,0}^{\rarr} &= \JLzp  + \NLzp d_{L,0}^{\rarr} +  \ALz  b_{L,0}^{\rarr}, \nonumber\\
b_{L,0}^{\rarr} &= -\left(1-t^2/2\right) c_{L,0}^{\rarr} +i t b_{R,-1}^{\rarr}, \nonumber\\
c_{L,0}^{\rarr} &= J_{L,0}^h + N_{L,0}^h b_{L,0}^{\rarr} + A_{L,0} d_{L,0}^{\rarr}, \nonumber\\
d_{L,0}^{\rarr} &= -\left(1-t^2/2\right) a_{L,0}^{\rarr} -i t d_{R,1}^{\rarr}, \nonumber\\[-6ex]
& & \qquad \qquad a_{R,-1}^{\rarr} &= - d_{R,-1}^{\rarr},  \nonumber\\
& & \qquad \qquad b_{R,-1}^{\rarr} &= A_{R,-1} a_{R,-1}^{\rarr} + N_{R,-1}^h c_{R,-1}^{\rarr},  \nonumber\\
& & \qquad \qquad c_{R,-1}^{\rarr} &= i t c_{L,0}^{\rarr} - b_{R,-1}^{\rarr},  \nonumber\\
& & \qquad \qquad d_{R,-1}^{\rarr} &= N_{R,-1}^p a_{R,-1}^{\rarr} + A_{R,-1} c_{R,-1}^{\rarr},  \nonumber \\[-6ex]
a_{L,-2}^{\rarr} &= N_{L,-2}^p d_{L,-2}^{\rarr} + A_{L,-2} b_{L,-2}^{\rarr}, \nonumber\\
b_{L,-2}^{\rarr} &=  -c_{L,-2}^{\rarr}, \nonumber\\
c_{L,-2}^{\rarr} &= N_{L,-2}^h b_{L,-2}^{\rarr} + A_{L,-2} d_{L,-2}^{\rarr}, \nonumber\\
d_{L,-2}^{\rarr} &= -a_{L,-2}^{\rarr},
\end{alignat}
where higher coefficients do not contribute to the perturbative current. Solving this set of equations, we obtain the following results for the dimensionless current density:
\begin{align}
T_{p,\rarr}^{p(0)}(E) &=(|a^{\rarr}_{L,0}|^2 - |d^{\rarr}_{L,0}|^2)\Theta(E + \mu_L), \nonumber\\*
&= t^2  \biggl[\frac{ \uL^2(E)\qpL(E)  }{k_{pL}(E)}\biggr] \, D_R(E_{1})\biggl[\frac{\qpR(E_1) \uR^2(E_1) + \qhR(E_1) \vR^2(E_1)}{k_{pR}(E_1)} \biggr]\Theta(E_1 + \mu_R)\Theta(E + \mu_L), \hspace{0.5 cm} \mathrm{and}\label{eq:Tp0p}
\end{align}
\begin{align}
T_{p,\rarr}^{h(0)}(E) &= (|c^{\rarr}_{L,0}|^2 - |b^{\rarr}_{L,0}|^2)\Theta(-E + \mu_L) \nonumber \\*
&= t^2 \biggl[\frac{ \vL^2(E)\qpL(E)  }{k_{hL}(E)}\biggr] D_R(E_{-1})\Biggl[\frac{\qhR(E_{-1}) \uR^2(E_{-1}) + \qpR(E_{-1}) \vR^2(E_{-1})}{k_{hR}(E_{-1})} \Biggr]\Theta(-E_{-1} + \mu_R) \Theta(-E + \mu_L).\label{eq:Th0p}
\end{align}
In going to the last lines of Eqs.~\eqref{eq:Tp0p} and~\eqref{eq:Th0p}, we have used the quasiparticle density of states
\begin{align}\label{eq:DR}
D_{j}(E) &= \frac{1}{u_j^2(E)-v_j^2(E)} \times\begin{cases}  \Theta\Bigl(|E| - \Delta_j \Bigr) \Theta\Bigl(E + \sqrt{\mu_j^2+\Delta_j^2} \Bigr), & \mu_j > 0,\\[1ex]
  \Theta\Bigl(E - \sqrt{\mu_j^2+\Delta_j^2} \Bigr),  & \mu_j \leq 0,
	\end{cases}\nonumber\\
	 &= \frac{|E|}{\sqrt{E^2-\Delta_j^2}} \times\begin{cases}  \Theta\Bigl(|E| - \Delta_j \Bigr) \Theta\Bigl(E + \sqrt{\mu_j^2+\Delta_j^2} \Bigr), & \mu_j > 0,\\
  \Theta\Bigl(E - \sqrt{\mu_j^2+\Delta_j^2} \Bigr)  & \mu_j \leq 0,
\end{cases}
\end{align}
for $j = R$. The dc current due to the quasiparticle injections from the left reservoir is
\begin{align}
I_{\mathrm{dc}}^{\rightarrow}(V) &= \frac{2}{h}t^2  \int_{-\infty}^{\infty} dE \, \Dl(E) \biggl\{ \biggl[\frac{\uL^2(E)\qpL(E)}{k_{pL}(E)}\biggr]\Theta(E + \mu_L) f(E)  \Dr(E+V)\Theta(E_{1} + \mu_R) \biggl[\frac{\qpR(E_1) \uR^2(E_1) + \qhR(E_1) \vR^2(E_1)}{k_{pR}(E_1)} \biggr]\nonumber\\
 &\qquad+\biggl[\frac{\vL^2(E)\qpL(E)}{k_{hL}(E)} \biggr] \Theta(-E + \mu_L) (1-f(E)) \Dr(E-V)\Theta(-E_{-1} + \mu_R) \Biggl[\frac{\qhR(E_{-1}) \uR^2(E_{-1}) + \qpR(E_{-1}) \vR^2(E_{-1})}{k_{hR}(E_{-1})} \Biggr]\biggl\}.
\end{align}
Changing the integration $E\to -E$ for the hole part and using $1-f(-E)=f(E)$, $D_{L/R}(-E) = D_{L/R}(E)$, and $k_{hL}(-E) = k_{pL}(E)$ gives the expression 
\begin{align}\label{eq:leftI}
I_{\mathrm{dc}}^{\rightarrow}(V) &= \frac{2}{h}t^2 \int_{-\infty}^{\infty} dE \Biggl\{\, D_{L}(E) \Theta(E + \mu_L) \biggl[\frac{\qpL(E) \uL^2(E) + \qhL(E) \vL^2(E)}{k_{pL}(E)} \biggr] \nonumber\\
&\hspace{3 cm}\times D_{R}(E+V)\Theta(E_{1} + \mu_R)  \biggl[\frac{\qpR(E_1) \uR^2(E_1) + \qhR(E_1) \vR^2(E_1)}{k_{pR}(E_1)} \biggr] f(E)\biggr\}\nonumber\\
&= \frac{2}{h}t^2 \int_{-\infty}^{\infty} dE \, \rho_{L}(E)   \rho_{R}(E+V)  f(E)
\end{align}
\end{widetext}
with the particle tunneling density of states ($j = L,R$)
\begin{align}
\rho_j(E) \equiv  D_{j}(E)  \biggl[\frac{q_{pj}(E) u_j^2(E) + q_{hj}(E) v_j^2(E)}{k_{pj}(E)} \biggr]\Theta(E + \mu_j) .
\end{align}
Note that in the Andreev approximation regime, $\rho_j(E) = D_j(E)$. Subtracting the current due to quasiparticle injections from the right reservoir $I_{\mathrm{dc}}^{\leftarrow}(V)$ gives the tunneling current~[Eq.\eqref{eq:tunnelling}]. 

%++++++++++++++++++++++++++++++++++++++++
\begin{figure*}[t]
\includegraphics[width=1.\linewidth]{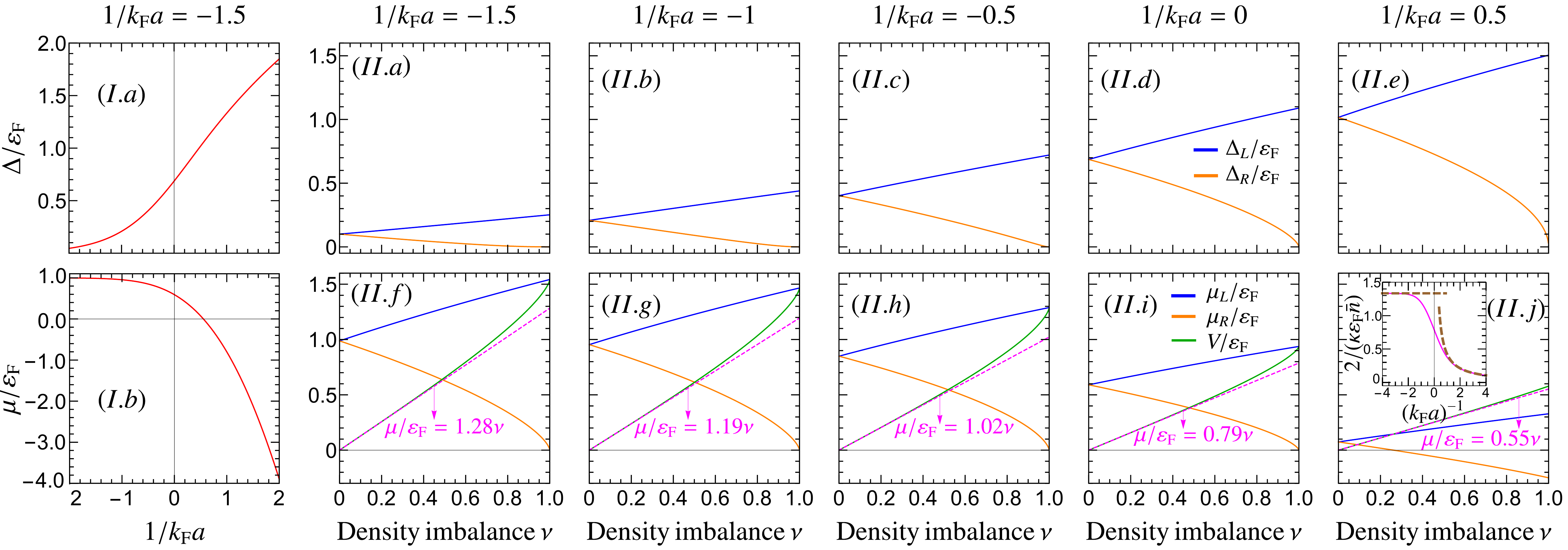}
\caption{ Left panel: (I.a) Mean-field gap and (I.b) chemical potential in units of the Fermi energy $\EF$ as a function of interaction strength $1/\kF a$ across the BCS-BEC crossover. [(II.a)--(II.e)] Pairing gaps of  the left ($\Delta_L$, blue line) and right ($\Delta_R$, orange) reservoirs. [(II.f)--(II.j)] Chemical potential of the left reservoir ($\mu_L$, blue), chemical potential of the right reservoir ($\mu_R$, orange) and bias voltage across the superfluid-normal-superfluid junction ($V = \mu_L - \mu_R$, green). All quantities are plotted in units of the Fermi energy $\EF$ as a function of density imbalance $\nu$ for different interaction strengths $1/\kF a$ along the BCS-BEC crossover. Inset (II.j): Plot of the proportionality constant between $V/\EF$ and $\nu$ as a function of interaction strength $1/\kF a$, where the proportionality constant is $4/3$ (dashed-brown line) deep in the BCS regime and is given by $3\kF a/(5\sqrt{2})$ (dashed-brown line) deep in the BEC regime.
}\label{fig:spectral}
\end{figure*}
%++++++++++++++++++++++++++++++++++++++++

\section{Bulk mean-field equations for the pairing gaps and chemical potentials}\label{sec:bulkmeanfield}

The main text presents results for the current across a Josephson junction along the BCS-BEC crossover. Here, we summarize results for the reservoir pairing gaps $\Deltal$ or $\Deltar$ and chemical potentials $\mu_L$ or $\mu_R$ obtained from a mean field calculation. These mean-field parameters are used in the main text for the reservoir parameters, although we emphasize that other parameters taken from more general many-body calculations can be used and the Landauer-B\"uttiker framework derived in this paper is independent of this choice. 

We describe a particular configuration in terms of the density imbalance~\eqref{eq:imbalance} between reservoirs as well as the interaction strength~\eqref{eq:interactionstrength}. Note that this definition implies \mbox{$n_{L/R} = (1\pm\nu) \bar{n}$} for the reservoir densities and thus, $\kFLR = (1 \pm \nu)^{1/3} \kF$ and $\EFLR = (1 \pm \nu)^{2/3} \EF$. In particular, results given in units of $\EFLR$ are expressed in terms of the common Fermi energy $\EF$ as
\begin{subequations}
\begin{align}
\frac{\mulr}{\EF} &= \bigl( 1 \pm \nu\bigr)^{2/3} \frac{\mulr}{\EFLR},  \\
\frac{\Deltalr}{\EF} &= \bigl( 1 \pm \nu\bigr)^{2/3} \frac{\Deltalr}{\EFLR}.
\end{align}
\end{subequations}

The mean-field equations are written in compact form as~\cite{eagles69,leggett80,bloch08,marini98}
\begin{align}\label{eq:meanfield}
\frac{\Delta}{\EF} &= \biggl[\frac{2}{3 I_2(y)}\biggr]^{2/3}, \nonumber\\
\frac{1}{\kF a} &= - \frac{2}{\pi} \biggl[\frac{2}{3 I_2(y)}\biggr]^{1/3} I_1(y) ,
\end{align}
where
\begin{align}
I_1(y) &= \int_0^\infty dx \, x^2 \biggl(\frac{1}{E_x} - \frac{1}{x^2}\biggr), \nonumber\\
I_2(y) &= \int_0^\infty dx \, x^2 \biggl(1 - \frac{\xi_x}{E_x}\biggr), \label{eq:intmeanfield}
\end{align}
and we define the dimensionless variables
\begin{align}
x^2 &= \frac{\hbar^2k^2}{2m\Delta}, \nonumber\\ 
y &= \frac{\mu}{\Delta}, \nonumber\\
\xi_x &= \frac{\xi_{\bf k}}{\Delta} = x^2 - y, \nonumber\\
E_x &= \frac{E_{\bf k}}{\Delta} = \sqrt{\xi_x^2 + 1}.
\end{align}

\begin{table*}[t]
\begin{tabular}{l c c c c }
 \multicolumn{4}{c}{(a) $h_n$: $V_n = \Delta_R/n$}          \\ \\
\hline
\hline
                            \multicolumn{1}{c}{$n$}& \multicolumn{1}{c}{$\nu$}  & \multicolumn{1}{c}{$V_n/\EF$} & \multicolumn{1}{c}{ $\Delta_R/\EF$}\\
\hline
\hline
1 &0.521 & 0.42 & 0.42  \\
2 &0.331 & 0.263 & 0.525 \\
3 &0.241 & 0.19 & 0.571\\
 4 &0.189 & 0.149 & 0.597 \\
 5 &0.155 & 0.123 &0.613 \\
6 &0.132 & 0.104 & 0.625 \\
 7 &0.115 & 0.09 & 0.633\\
8 &0.101 & 0.08 & 0.639 \\
\hline
\end{tabular} \qquad\qquad
%\\
\vspace{0.5 cm}
\begin{tabular}{l c c c c }
 \multicolumn{5}{c}{(b) $i_n$: $V_n = (\mu_L+\Delta_R)/(2n-1)$}          \\ \\
\hline
\hline
                            \multicolumn{1}{c}{$n$}& \multicolumn{1}{c}{$\nu$}  & \multicolumn{1}{c}{$V_n/\EF$}   & \multicolumn{1}{c}{ $\mu_L/\EF$} & \multicolumn{1}{c}{ $\Delta_R/\EF$}\\
\hline
\hline
2 &0.5 & 0.402 & 0.774 & 0.432\\
3 &0.313 & 0.249 & 0.708 & 0.534\\
 4 &0.227 & 0.179 & 0.677 & 0.578\\
 5 &0.178 & 0.14 &0.659 & 0.603 \\
6 &0.146 & 0.115 & 0.647 & 0.618\\
 7 &0.124 & 0.097 & 0.638 & 0.629\\
8 &0.107 & 0.085 & 0.632 & 0.636 \\
\hline \\
\end{tabular}\qquad\qquad
%\\
\vspace{0.5 cm}
\begin{tabular}{l c c c c }
 \multicolumn{4}{c}{(c) $j_n$: $V_n = \mu_R/n$}          \\ \\
\hline
\hline
                            \multicolumn{1}{c}{$n$}& \multicolumn{1}{c}{$\nu$}  & \multicolumn{1}{c}{$V_n/\EF$}   &  \multicolumn{1}{c}{ $\mu_R/\EF$}\\
\hline
\hline
1 &0.478 & 0.383 & 0.383\\
2 &0.295 & 0.234 &  0.468\\
 3 &0.212 & 0.168 &  0.504\\
 4 &0.166 & 0.131 &0.523 \\
5 &0.136 & 0.107 & 0.536\\
 6 &0.115 & 0.091 &  0.544\\
\hline \\ \\
\end{tabular}\\[-5ex]
\caption{
Density imbalance $\nu$ and voltage $V_n$ at which the $n$-th order of the subharmonic gap structure occurs at unitarity. Shown also are the values of the chemical potentials ($\mu_L,\mu_R$) and pairing potentials ($\Deltal, \Deltar$) of the left ($L$) and right ($R$) reservoirs. A plot of the chemical potentials, pairing potentials and voltages as a function of density imbalance at unitarity can also be found in Figs.~\ref{fig:spectral}(II.d) and~\ref{fig:spectral}(II.i).
}\label{table:SGS}
\end{table*}

Both integrals are evaluated in terms of complete elliptic integrals of the first and second kind. Figures~\ref{fig:spectral}(I.a) and~\ref{fig:spectral}(I.b) show the mean-field result for the gap and chemical potential of a single reservoir as a function of scattering length across the BCS-BEC crossover. As shown in the figure, the condition \mbox{$\Delta \ll \mu$} corresponding to the Andreev approximation is satisfied only in the strict BCS limit [$1/(k_Fa) \ll 0$]. In the unitary limit, the gap and chemical potential are of comparable magnitude, and on the BEC-side the chemical potential even turns negative. For two reservoirs, Fig.~\ref{fig:spectral} shows the mean-field gap $\Deltalr$ [upper panel: (II.a)--(II.e)] and chemical potential $\mulr$ [lower panel: (II.f)--(II.j)] as a function of density imbalance $\nu$ for five different interaction strengths \mbox{$(\kF a)^{-1} = -1.5,-1,-0.5,0,$} and $0.5$, corresponding to the parameter values in Fig.~\ref{fig:3} of the main text. The green line indicates the chemical potential difference or bias voltage \mbox{$V=\mu_L-\mu_R$}. We see that the Andreev approximation (\mbox{$\Delta \ll \mu$}) breaks down even for moderate deviations from the BCS regime at $(\kF a)^{-1} = -1$ [Figs.~\ref{fig:spectral}(II.b) and~\ref{fig:spectral}(II.g)].

The bias voltage [shown as green curves in Figs.~\ref{fig:spectral}(II.f)--\ref{fig:spectral}(II.j)] is linearly proportional to the density imbalance for $\nu \lesssim 0.5$, with a constant given by $2/(\kappa \EF \bar{n})$, according to Eq.~\eqref{eq:VEF}. The inset of Fig.~\ref{fig:spectral}(II.j) shows that the proportionality constant decreases from a value of $4/3$ deep in the BCS regime [which follows directly from the chemical potential $\mu = \hbar^2 (3\pi \bar{n})^{2/3}/(2m)$ of a free Fermi gas], assumes a value of $\approx 0.79$ at unitarity ($1/\kF a = 0$), and vanishes as $3\kF a/(5\sqrt{2})$ deep in the BEC regime.

\section{Subharmonic gap structure}\label{sec:SGS}

In this appendix, we list in Table~\ref{table:SGS} the numerical values for the positions of the subharmonic gap structure in the current-voltage characteristic calculated at unitarity shown in Fig.~\ref{fig:SGS} of the main text.

\bibliography{bib_sns}

\end{document}